\RequirePackage{fix-cm}
\documentclass[smallextended]{svjour3}       
\smartqed  
\usepackage{graphicx}
\usepackage{url}
\usepackage{amsmath}
\usepackage{comment}

\begin{document}

\title{Reconstruction of Interbank Network using Ridge Entropy Maximization Model}



\author{Yuichi Ikeda \and Hidetoshi Takeda 
}


\institute{Y. Ikeda* \at
              Graduate School of Advanced Integrated Studies in Human Survivability, Kyoto University, Kyoto 606-8306, JAPAN \\
              Tel.: +81-75-762-2102\\
              \email{ikeda.yuichi.2w@kyoto-u.ac.jp}  \\
              * to whom correspondence should be addressed. \\
           \and
           H. Takeda \at
              Graduate School of Advanced Integrated Studies in Human Survivability, Kyoto University, Kyoto 606-8306, JAPAN \\
			present affiliation: College of International Relations, Nihon University, Shizuoka 411-8555, JAPAN \\
              \email{takeda.hidetoshi@nihon-u.ac.jp}   \\
}

\date{Received: date / Accepted: date}

\maketitle

\begin{abstract}

We develop a network reconstruction model based on entropy maximization considering the sparsity of networks.
We reconstruct the interbank network in Japan from financial data in individual banks' balance sheets using the developed reconstruction model from 2000 to 2016.
The observed sparsity of the interbank network is successfully reproduced.
We examine the characteristics of the reconstructed interbank network by calculating important network attributes.
We obtain the following characteristics, which are consistent with the previously known stylized facts.
Although we do not introduce the mechanism to generate the core and peripheral structure, we impose the constraints to consider the sparsity that is no transactions within the same bank category except for major commercial banks, the core and peripheral structure has spontaneously emerged.
We identify major nodes in each community using the value of PageRank and degree to examine the changing role of each bank category.
The observed changing role of banks is considered a result of the quantitative and qualitative monetary easing policy started by the Bank of Japan in April 2013.

\keywords{Network reconstruction \and Interbank network \and Ridge entropy maximization}
\end{abstract}

\section{Introduction}
\label{intro}

Policymakers had traditionally focused on ensuring the financial soundness of individual financial institutions, especially deposit-taking institutions. Such an approach is known as the microprudential policy. However, the global financial crisis made it clear that keeping individual financial institutions sound was not enough. Based on the experience of the crisis, policymakers of many countries understood that it was very important to analyze and assess risks in the entire financial system and to take adequate measures to limit systemic risk. This is the background of the emergence of the macroprudential policy \cite{BOJ2011a} \cite{IMF-FSB-BIS2016} \cite{Sato2014}.
Since the global financial crisis, many countries are expanding their toolkits for more systemic approaches to financial regulation and supervision based on the macroprudential framework.

From the understandings of the necessity of the macroprudential policy, we studied channels of distress propagation from the financial sector to the real economy through the supply chain network in Japan from 1980 to 2015, using a spin dynamics model \cite{Ikeda2018}.
This study was conducted based on the related econophysics studies \cite{Ikeda2014}, \cite{Ikeda2015}, \cite{Ikeda2016}, \cite{Aoyama2017}.
However, the interbank network was not considered in this paper because data of interbank transactions are not available to the public.
In the macroprudential policy framework, particular attention is paid to the consequences of the interconnectedness among financial institutions, financial markets, and other components of the financial system. Also, feedback loops between real economies and financial systems are checked carefully.

The goal of this paper is to reconstruct the adjacency matrix of the interbank network.
We develop a network reconstruction model based on entropy maximization considering the sparsity of the network. Here the reconstruction is to estimate the network's adjacency matrix from the node's local information.
We reconstruct the interbank network in Japan from financial data in individual banks' balance sheets using the developed reconstruction model. Here the interbank money market works for short-term lending and borrowing between banks.
We examine the validity of the reconstructed interbank network by analyzing the interbank network's characteristics.

This paper is organized as follows.
In section \ref{sec:2}, preceding studies for interbank market and network reconstruction are explained.
In section \ref{sec:3}, the ridge entropy maximization model is proposed as a reconstruction model considering the sparsity of the real-world networks.
In section \ref{sec:4}, data used to reconstruct the interbank network is explained.
In section \ref{sec:5}, the reconstruction results are shown, and the accuracy and characteristics of the reconstructed interbank network are discussed.
Finally, section \ref{sec:6} concludes the paper.

\section{Literature Survey}
\label{sec:2}

\subsection{Interbank Market}
\label{sec:2-1}

Interbank markets are a type of money markets in which only financial institutions, such as banks, securities companies, and money dealers, participate and lend/borrow funds maturing within a year. Many central banks have set the overnight interbank interest rate as a target rate (policy interest rate). Therefore, interbank markets are very important for central bank operations for their monetary policy.
The global financial crisis had serious impacts on interbank markets. The environment surrounding interbank markets has changed significantly since the crisis, but interbank markets continue to play important roles as a center for fund transactions among financial institutions.
Many countries work to make their interbank markets functional, for example, by providing settlement services through market participants’ current deposit accounts at central banks and launching computer systems for funds transfers/settlements as market infrastructures. The following explains the mechanisms for facilitating the functions of interbank markets.

\paragraph{Current Deposit Accounts at Central Banks}

Central banks provide safe and convenient settlement assets in the form of deposits in current accounts that financial institutions hold at central banks.

Current account deposits at central banks serve three significant roles:
(1) Payment instrument for transactions among financial institutions, the Bank, and the government;
(2) Cash reserves for financial institutions to pay individuals and firms; and
(3) Reserves of financial institutions subject to the reserve requirement system.
Most central banks - more than 90 \% according to Gray \cite{Gray2011} - have adopted reserve requirement systems (RRS). Under the RRS, deposit-taking institutions are required to put at least a particular portion of their deposits received from their depositors at their central Bank current accounts. The portion is determined by the Central Bank of the country. The minimum amount that deposit-taking institutions must have in their current account at central banks is the “minimum reserve requirements”.

Gray \cite{Gray2011} states three primary purposes of the RRS: prudential, monetary control, and liquidity management.
First, the RRS serves as a means of prudential by ensuring that banks hold a certain proportion of high-quality, liquid assets.
Second, the RRS serves as a means of monetary control. The uses of reserve requirements are usually described in terms of two channels: the money multiplier and the impact of reserve requirement on interest rate spreads.
Last, the RRS serves as a means of liquidity management through averaging reserve balances, decreasing surplus reserve balances.

The RRS was initially introduced for prudential purposes. Later, in the 1930s, changes in reserve requirement ratios began to be used as a means of monetary policy \cite{Monnet2019}. Currently, central banks of developed countries with advanced financial markets conduct their monetary policies through operations in their short-term financial markets, and reserve requirement ratios are not used as a means of monetary policy. For instance, the Bank of Japan has not changed the reserve requirement ratio since October 1991.

\paragraph{Unconventional Monetary Policy and Reserves}

In the 2000s (especially after the global financial crisis), central banks of developed countries have implemented so-called unconventional monetary policies and have greatly expanded their balance sheets. As a result, these central banks now have huge excess reserve balances (reserves that exceed the required level). Under such a circumstance, by paying interest on excess reserve balances as a practical lower bound on the interbank market interest rate, the central banks started making use of such reserve balances as a tool for market operations and also for keeping market functions \cite{FRB2016}, \cite{BOJ2008}.

\paragraph{Fund Settlement System for Interbank Transactions}

Many central banks provide online fund settlement services as infrastructures for efficient and secure fund settlements among financial institutions. Such services are provided through large computer systems, such as Bank of Japan Financial Network System (BoJ Net), Fedwire Funds (US), and TARGET2 (Euro area). Transactions in interbank markets are settled by fund transfer between two accounts held at the central bank. The smooth operation of these funds settlement systems is a prerequisite for the full functioning of the interbank markets.

Fund settlement services provided by central banks have common features. For instance, the fund settlements through their settlement systems are final. Also, transactions are settled based on the “Real Time Gross Settlement (RTGS)” method to reduce settlement risk.
The introduction of the RTGS system comes from considerations to limit systemic risk. In the deferred net settlement (DNS) system that had been common in the past, a fail of settlement at the designated time by a participant may trigger consecutive fails in settlements. This comes from the fact that a participant often makes use of the funds to be received from other participants for their own payments. Thus, there was a risk that failure of payment by a participant might stop whole settlements among many financial institutions. This would be a form of the materialization of systemic risk. For this reason, many countries switched from the DNS system to the RTGS system. With the RTGS method, funds are settled immediately for each order based on instructions from the financial institution. Therefore, the direct impact of default is limited to its instructed counterparty \cite{BOJ2011b}. According to a World Bank survey, as of 2016, 103 out of 113 countries (91\%) have adopted the RTGS fund settlement system \cite{WorldBank2018}.

\subsection{Network Reconstruction}
\label{sec:2-2}

The network reconstruction has been an actively studied topic in network science and a large number of works have been published during the year \cite{Anand2018}, \cite{Anand2014}, \cite{Wilson1967}, \cite{Duenas2013}, \cite{Drehmann2013}, \cite{Mastrandrea2014}, \cite{Cimini2015a}, \cite{Cimini2015b}, \cite{Ikeda2017}, \cite{Squartini2018}, \cite{Ramadiah2020}.
The following is a brief desciption of models closely related to the model described in section \ref{sec:3}.

\paragraph{MaxEnt model}

The MaxEnt model maximizes the entropy $S(t_{ij})=-\sum t_{ij} \ln t_{ij}$ by changing the amount of outstanding loans made by bank $i$ to bank $j$, $t_{ij}$ under the following constraints  \cite{Wells2004}, \cite{Upper2011}:
\begin{equation}
 s_{i}^{out} = \sum_j t_{ij},  
\label{Const1}
\end{equation}
\begin{equation}
 s_{j}^{in} = \sum_i t_{ij},
\label{Const2}
\end{equation}
\begin{equation}
 G = \sum_i s_{i}^{out} = \sum_j s_{j}^{in},  
\label{Const3}
\end{equation}
where the amount of aggregated loans owned by bank $i$, $s_{i}^{out}$ and the amount of aggregated borrowing of bank $j$, $s_{j}^{in}$ are assumed to be given.
The analytical solution is easily obtained as
\begin{equation}
 t_{ij}^{ME} = \frac{s_{i}^{out} s_{j}^{in}}{G}.  
\label{MaxEnt4}
\end{equation}
It is, however, noted that the solution in Eq. (\ref{MaxEnt4}) gives a fully connected network, although the real-world networks are often known as sparse networks.
We note that $t_{ij}^{ME}$ in Eq. (\ref{MaxEnt4}) is used as the null model in the definition of modularity $M$ used in the community analysis of networks \cite{Newman2004}.

\paragraph{Other major models} 

The other significant models were systematically surveyed in Appendix B of the literature \cite{Anand2018}.
Here, only a summary of the selected models is given.

The Minimum Density model produces networks preserving the characteristic of interbank networks \cite{Anand2014}. The authors note that interbank activity is based on sparse relationships in interbank networks. This model is formulated as a constrained optimization problem. The objective function is the total cost of interbank transactions. As constraints, aggregated interbank assets and aggregated liabilities of each bank are given. This problem, however, is computationally expensive to solve. The authors proposed a heuristic to solve this problem.

A fitness model postulates that the probability of a bank acquiring links is proportional to its fitness \cite{Musmeci2013}.
The probability for a link between two banks i and j is formulated using the fitness $f_i$ $(j)$ of bank $i$ $(j)$ with parameter z.
First, the parameter z is estimated from the aggregate lending and borrowing constraints of banks. Second, a series of adjacency matrices are sampled using the probabilities for the link. Finally, the exposures are determined using the standard maximum entropy method.

Another fitness model is presented related to Musmeci's model but with some critical differences \cite{Cimini2015a} \cite{Cimini2015b}. First, both methods generate adjacency matrices from so-called fitness models. However, in Musmeci's model, the matrices are undirected, while Cimini's model is directed. Second, for assigning the exposures, Musmeci's model utilizes the maximum entropy method. While in Cimini's model, the exposure assignment also follows a fitness model.
The aggregate exposure constraints are satisfied only in an average over a large number of reconstructed networks.

\section{Ridge Entropy Maximization Model}
\label{sec:3}

\paragraph{Convex Optimization} 
\label{Convex}

Configuration entropy $S$ is written using bilateral transaction $t_{ij}$ between banks $i$ and $j$ as follows,
\begin{equation}
 S = \log \frac{ \left( \sum_{ij} t_{ij} \right) ! }{ \prod_{ij} t_{ij} ! } \approx \left( \sum_{ij} t_{ij} \right) \log \left( \sum_{ij} t_{ij} \right) - \sum_{ij} t_{ij} \log t_{ij}.
\label{entropy1}
\end{equation}
Here an approximation is applied to factorial $!$ using Stirling's formula.   
The first term of R.H.S. of Eq.(\ref{entropy1}) does not change the value of $S$ by changing $t_{ij}$ because $\sum_{ij} t_{ij}$ is constant. Consequently, we have a convex objective function:
\begin{equation}
 S = - \sum_{ij} t_{ij} \log t_{ij}.
\label{entropy2}
\end{equation}
Entropy $S$ is to be maximized under the constraints given by Eqs. (\ref{Const1}) - (\ref{Const3}),
where $s_{i}^{out}$ and $s_{j}^{in}$ correspond to the local information given for each node.

\paragraph{Sparsity of Network} 
\label{Sparse}

The accuracy of the reconstruction will be improved using the sparsity of the interbank network.
The sparsity is characterized by the skewness of the observed in-degree and out-degree distributions. This means that a limited fraction of nodes have a large number of links and most nodes have a small number of links and consequently the adjacency matrix of international trade is sparse.

To take into account the sparsity,
the objective function in Eq. (\ref{entropy2}) is modified by applying the concept of Lasso (least absolute shrinkage and selection operator)  or ridge regularization \cite{Tibshirani1996}  \cite{Breiman1995} \cite{Hastie2008} to our convex optimization problem.

\paragraph{Lasso or Ridge Regularizations}

In the case of Lasso or L1 regularization, our problem is formulated as the maximization of objective function $z$: 
\begin{equation}
 z(t_{ij}) = S - \sum_{ij} \left| t_{ij} \right| = - \sum_{ij} t_{ij} \log t_{ij} - \beta \sum_{ij} \left| t_{ij} \right|
\label{Lasso}
\end{equation}
with local constraints. Here the second term of R.H.S. of Eq. (\ref{Lasso}) is L1 regularization. $\beta$ is a control parameter. However the L1 regularization term $\sum_{ij} \left| t_{ij} \right|$ is constant in our problem. Therefore, we need a variant of the Lasso concept, e.g. ridge or L2 regularization.

By considering this fact, our problem is reformulated as the maximization of objective function $z$:
\begin{equation}
 z(t_{ij}) = S - \sum_{ij} t_{ij}^2 = - \sum_{ij} t_{ij} \log t_{ij} - \beta \sum_{ij} t_{ij}^2
\label{Ridge}
\end{equation}
with local constraints. Here the second term of R.H.S. of Eq. (\ref{Ridge}) is L2 regularization.

In the Lasso regularization, many transaction $t_{ij}$ become zero and only small number of variables are finite values.
For this reason, Lasso regression is called sparse modeling.
On the other hand, in the ridge regularization, relatively small values are obtained for many transaction $t_{ij}$.

\begin{figure*}
  \includegraphics[width=0.35\textwidth]{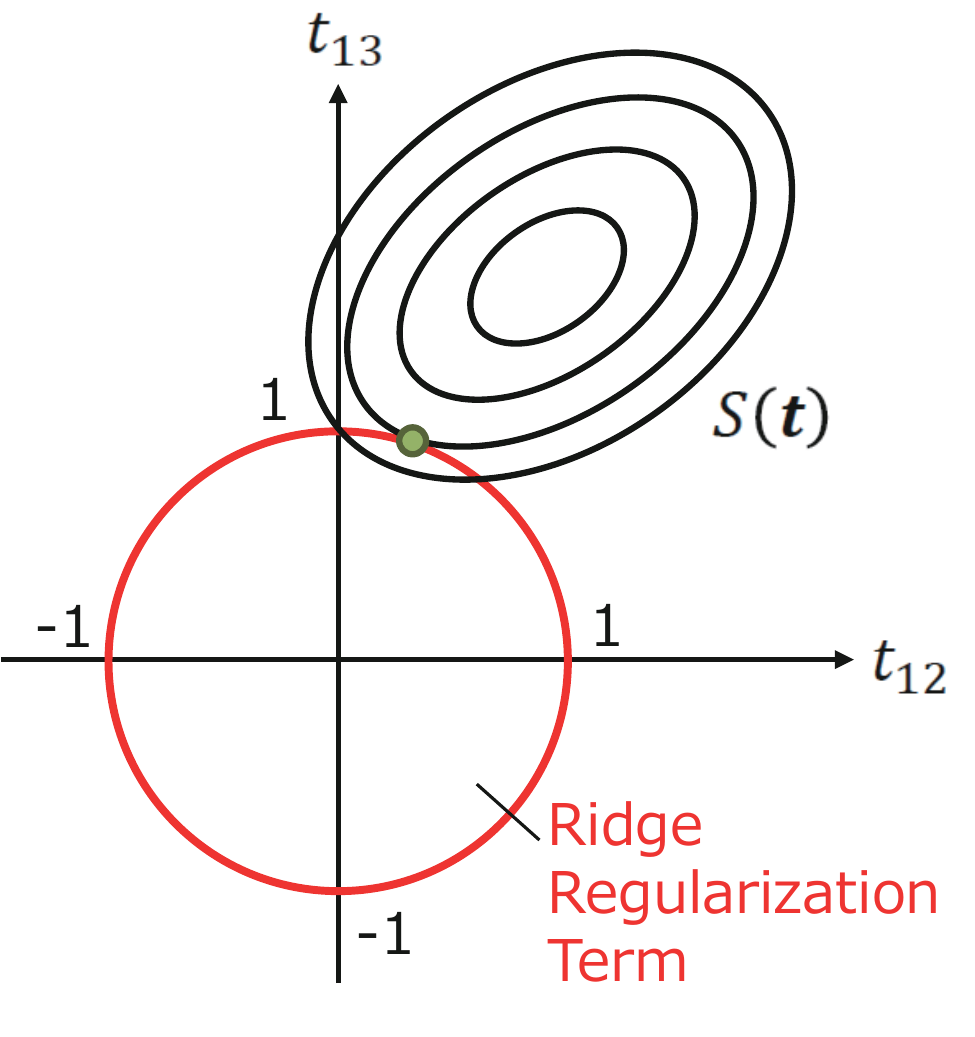}
  \includegraphics[width=0.35\textwidth]{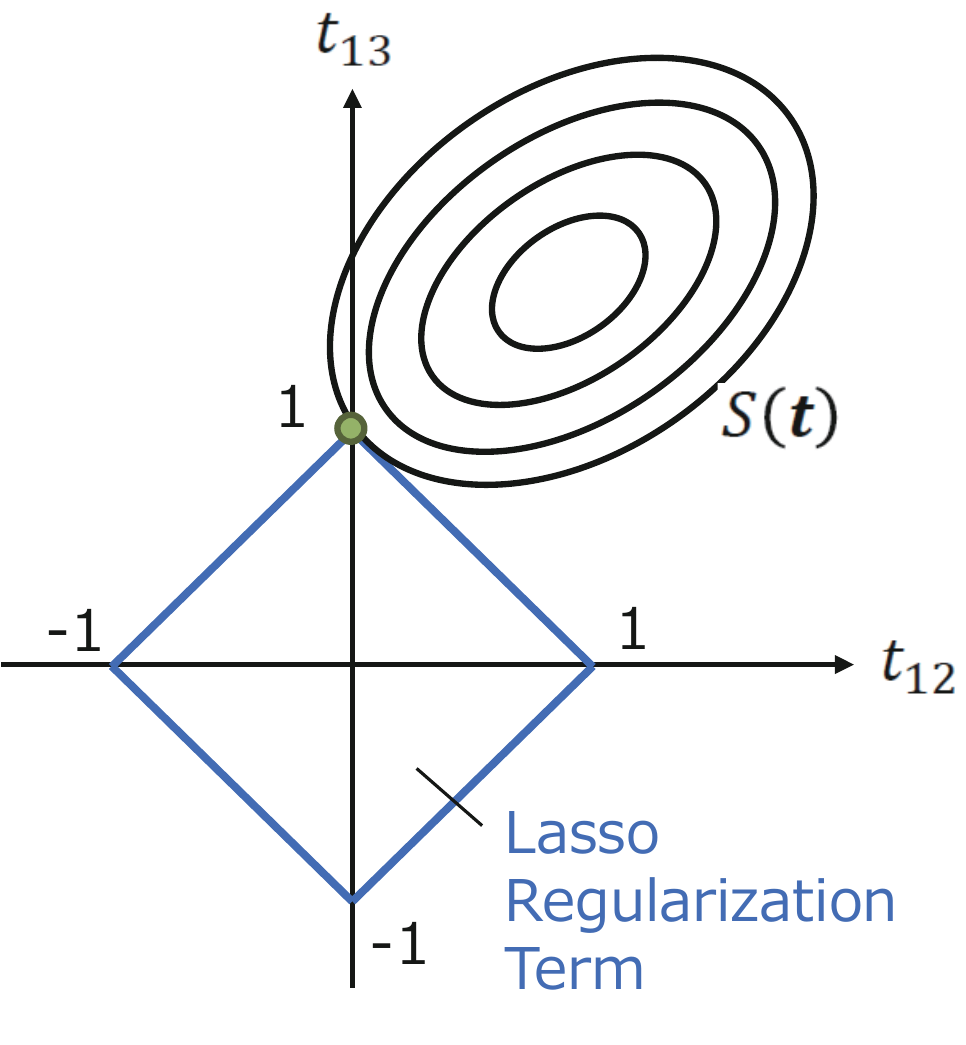}
\caption{The Ridge regression yields small overall values for the variables $t_{ij}$, except for significant explanatory variables. On the other hand, Lasso regression yields many variables $t_{ij}$ that are zero and only some variables that are different from zero. For this reason, Lasso regression is called sparse modeling.}
\label{fig:1}       
\end{figure*}

\paragraph{Reconstruction Model considering Sparsity}

In the theory of thermodynamics, the equilibrium of a system is obtained by minimizing thermodynamic potential $F$:
\begin{equation}
 F = E - TS
\label{FreeEnergy}
\end{equation}
where $E$, $T$, and $S$ are internal energy, temperature, and entropy, respectively. Eq. (\ref{FreeEnergy}) is rewritten as a maximization problem as follows,
\begin{equation}
 z \equiv -\frac{1}{T} F = S - \frac{1}{T} E.
\label{ObjtFcun}
\end{equation}
We note that Eq. (\ref{ObjtFcun}) has the same structure to Eq. (\ref{Ridge}). Thus we interpret the meaning of control parameter $\beta$ and L2 regularization term as inverse temperature $1/T$ and internal energy $E$, respectively.

We obtain the ridge entropy maximization model by replacing interbank transaction $t_{ij}$ by $p_{ij}={t_{ij}}/{\sum_{ij} t_{ij}}$ \cite{Ikeda2018b}: 
\begin{equation}
\renewcommand{\arraystretch}{2.3}
\begin{array}{ll}
\mbox{maximize   } & z (p_{ij}) = - \sum_{ij} p_{ij} \log p_{ij} - \beta \sum_{ij} p_{ij}^2 \\ 
\mbox{subject to   } & \sum_{ij} p_{ij} = 1 \\
                       & p_{ij} \geq 0 \\ 
                       & \frac{s_{i}^{out}}{G} = \sum_j p_{ij} \\ 
                       & \frac{s_{j}^{in}}{G} = \sum_i p_{ij} \\
                       & G = \sum_{ij} t_{ij} \\
\end{array}
\label{ConvexFormulation}
\end{equation}

Based on  the ridge entropy maximization model formulated in Eqs. (\ref{ConvexFormulation}), this paper proposes the following new constraints that can be applied to a variety of real-world problems.
First, let us consider a network reconstruction among banks that form a subset of all financial institutions.
In this case, the amount of aggregate loan owned by the banks $\sum_i s_{i}^{out}$ is not equal to the amount of aggregate borrwoing of the banks $\sum_j s_{j}^{in}$. 
Therefore, node ``other'' is needto be added as a slack variable in the ridge entropy maximization model formulated in Eqs. (\ref{ConvexFormulation}) to satisfy $\sum_{ij} p_{ij} = 1$.

In addition, consider the case where the banks to be analyzed can be divided into several categories and the total amount of transactions among a particular set of categories is known. This includes cases where the total amount of transactions within a particular category is known. In these cases, we can add the following constraints,
\begin{equation}
 \sum_{i \in g_l} \sum_{j \in g_l'}p_{ij} =\frac{Q_{ll'}}{G},
\label{GroupC}
\end{equation}
where $g_l$ is the $l \mathchar`-{\text{th}}$ group and $Q_{ll'}$ is the total amount of transactions between the the $l\mathchar`-{\text{th}}$ category and the $l'\mathchar`-{\text{th}}$ category.
As a special case of Eq. (\ref{GroupC}), if no transaction among a particular set of categories is made $Q_{ll'}=0$, we impose a constraint the following conditions,
\begin{equation}
 p_{ij} =0 \; \; \; \; \; \;  (i \in g_l, j \in g_l').
\label{GroupC2}
\end{equation}

\section{Financial and Economic Data}
\label{sec:4}

The Japanese banks are categorized as the major commercial bank, the trust bank, the leading regional bank, and the second-tier regional bank.
The major commercial banks are financial institutions that provide financial intermediary services such as underwriting government bonds through international financial markets. They are represented by Mitsubishi UFJ Bank, Sumitomo Mitsui Banking Corporation, Mizuho Bank, and Resona Bank. The trust banks are financial institutions mainly engaged in the trust business. They include Mitsubishi UFJ Trust and Banking Corporation and Norinchukin Trust and Banking Corporation. The leading regional banks are the largest financial institutions in each prefecture and greatly influence the local economy. Approximately, Sixty banks fall under this category. The second-tier regional banks are financial institutions in each prefecture, but they are smaller than the leading regional banks, and some of them are unlisted. More than thirty banks fall under this category.

In this section, we explain the data used in the reconstruction of the interbank network.

\paragraph{Lending and Borrowing in Interbank Market}

Imakubo and Soemjima studied the transaction in Japanese interbank market.
The core parts of Table 4 in paper \cite{Imakubo2010}  are reproduced in Table \ref{table:CallLoanMoney}.

This study shows only a small amount of transaction is observed among the trust banks, the leading regional banks, and the second-tier regional banks at the end of December, 2005.
(1) Lending from the trust banks to the trust banks, the regional banks, and the 2nd tier regional banks are $0.3 \times 10^{12}$, $0.6 \times 10^{12}$, and $0.0 \times 10^{12}$ in JPY, respectively.
(2) Lending from the regional banks to the trust banks, the regional banks, and the 2nd tier regional banks are $1.4 \times 10^{12}$, $1.0 \times 10^{12}$, and $0.0 \times 10^{12}$ in JPY, respectively.
(3) Lending from the 2nd-tier regional banks to the trust banks, the regional banks, and the 2nd tier regional banks are $0.1 \times 10^{12}$, $0.1 \times 10^{12}$, and $0.0 \times 10^{12}$ in JPY, respectively.
Especially, market transactions within the same bank category are practically non-existent except for major commercial banks.
We consider this sparsity to reconstruct the interbank network in Japan by imposing the constraints in Eq. (\ref{GroupC2}).
\begin{table}[hbtp]
  \caption{Interbank Market in December 2005 ( $10^{12}$ JPY, Table 4 in paper \cite{Imakubo2010}) }
  \label{table:CallLoanMoney}
  \centering
  \begin{tabular}{lccccc}
    \hline
    Lender  &  Major  & Rrust & Leading regional & 2nd tier regional & Total \\
    \hline \hline
    Major bank        & 7.0 & 2.2 & 2.4 & 0.0 & 11.6 \\
    Trust bank        & 4.2 & 0.3 & 0.6 & 0.0 & 5.1 \\
    Leading regional & 7.4 & 1.4 & 1.0 & 0.0 & 9.8 \\
    2nd tier regional & 3.2 & 0.1 & 0.1 & 0.0 & 3.4 \\
    Total                & 21.8 & 4.0 & 4.1 & 0.0 & 29.9 \\
    \hline
  \end{tabular}
\end{table}

\paragraph{Call Loan and Call Money in Balance Sheet}

\begin{figure*}
  \includegraphics[width=0.5\textwidth]{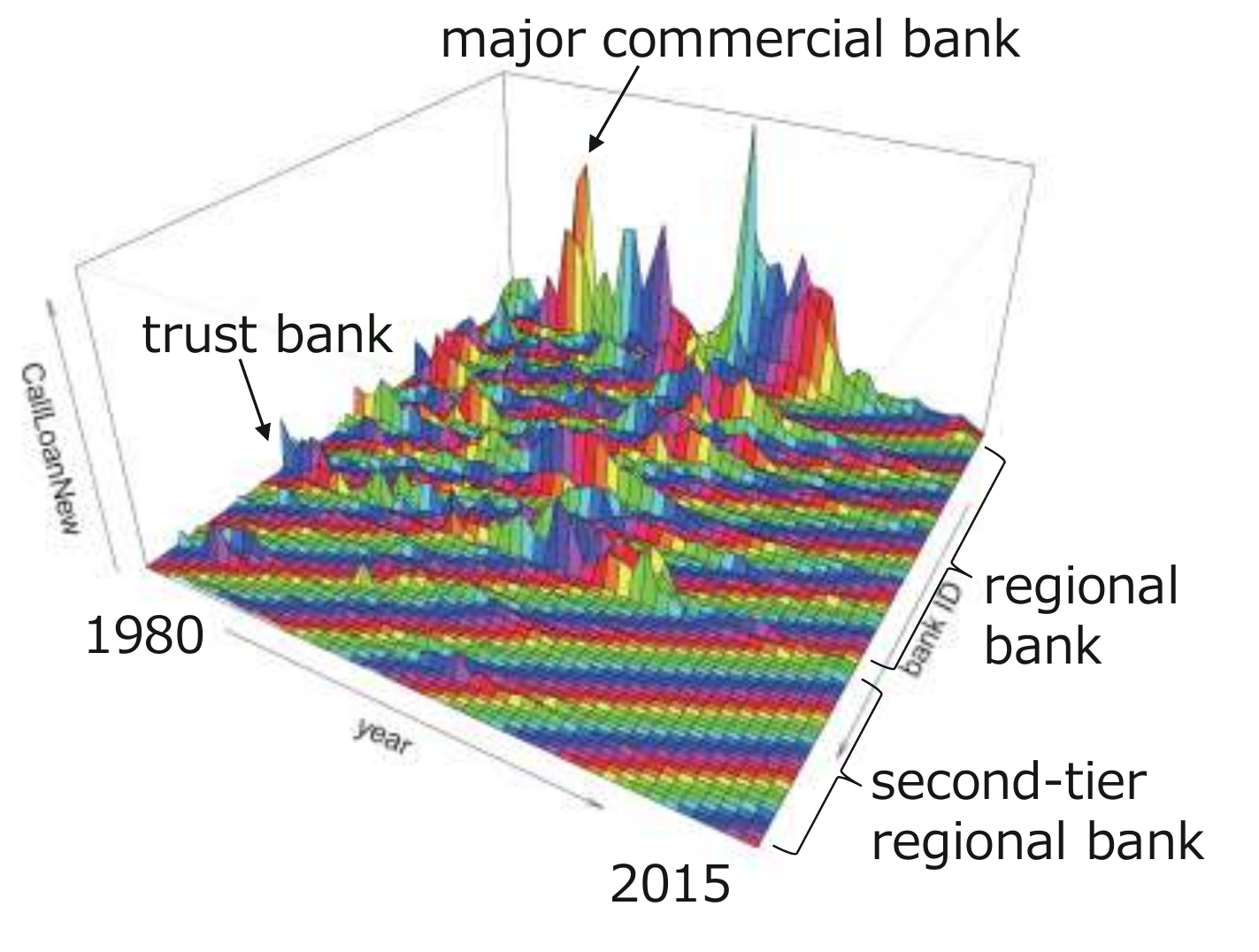}
  \includegraphics[width=0.5\textwidth]{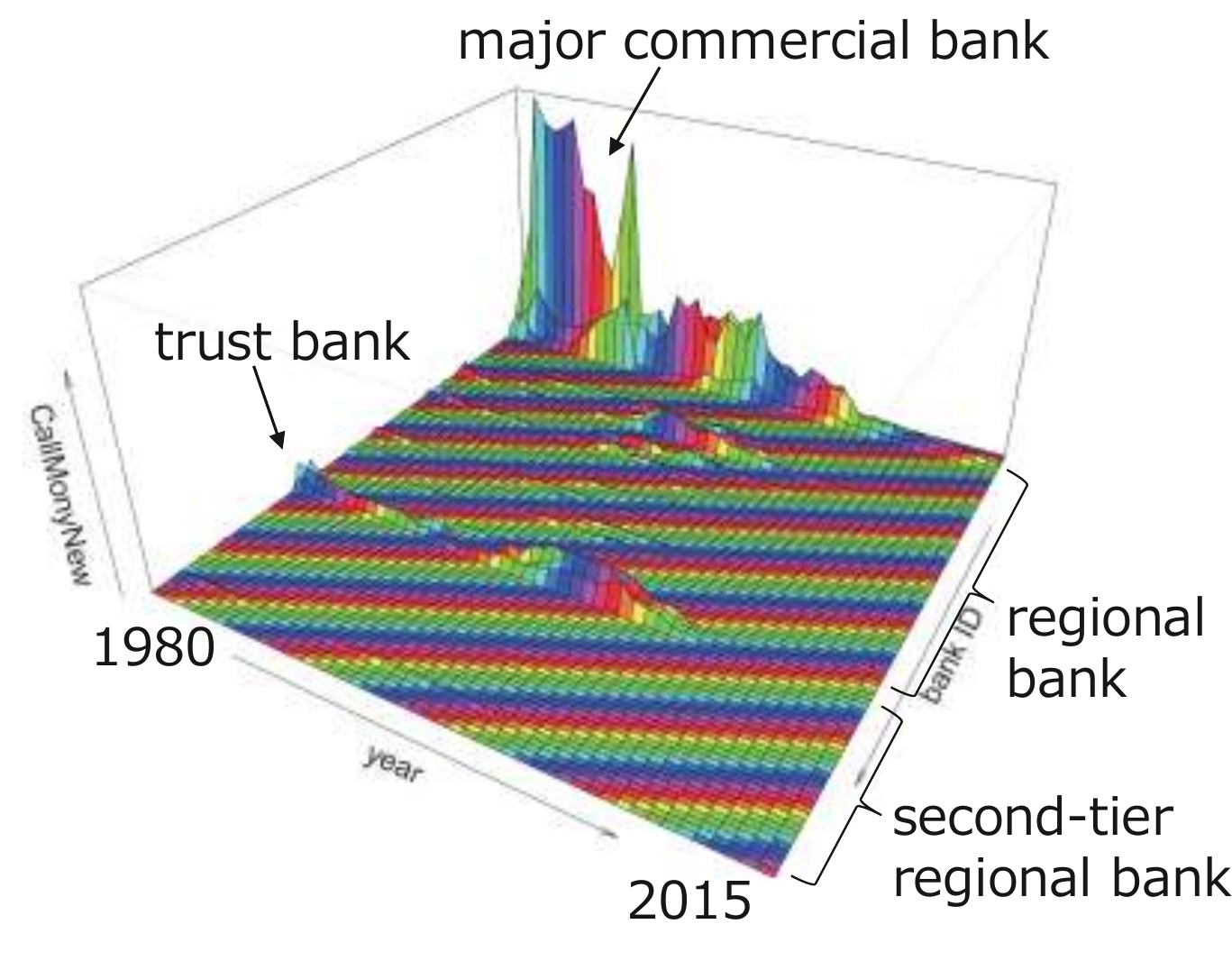}
\caption{Temporal change of call loan (left panel) and call money (right panel) for each bank. 
The leading regional banks lent money to the major commercial banks and the trust banks. The amount of both call loan and call money has decreased since the early 2000s. This coincides with the recent increase of purchasing government bond.
}
\label{fig:TemporalChange}       
\end{figure*}

Call loan (lending) and call money (borrowing) of 98 banks are recorded in balance sheets.
Temporal change of call loan and call money for each bank are shown in Fig. \ref{fig:TemporalChange}.
The leading regional banks have a large amount of call loans, and the major commercial banks and the trust banks have large amount of call money.
This implies that
the leading regional banks lend money to the major commercial banks and the trust banks. It is, however, noted that the amount of both call loan and call money has decreased since the early 2000s. This coincides with the recent increase in purchasing government bonds.

\begin{figure*}
  \includegraphics[width=0.4\textwidth]{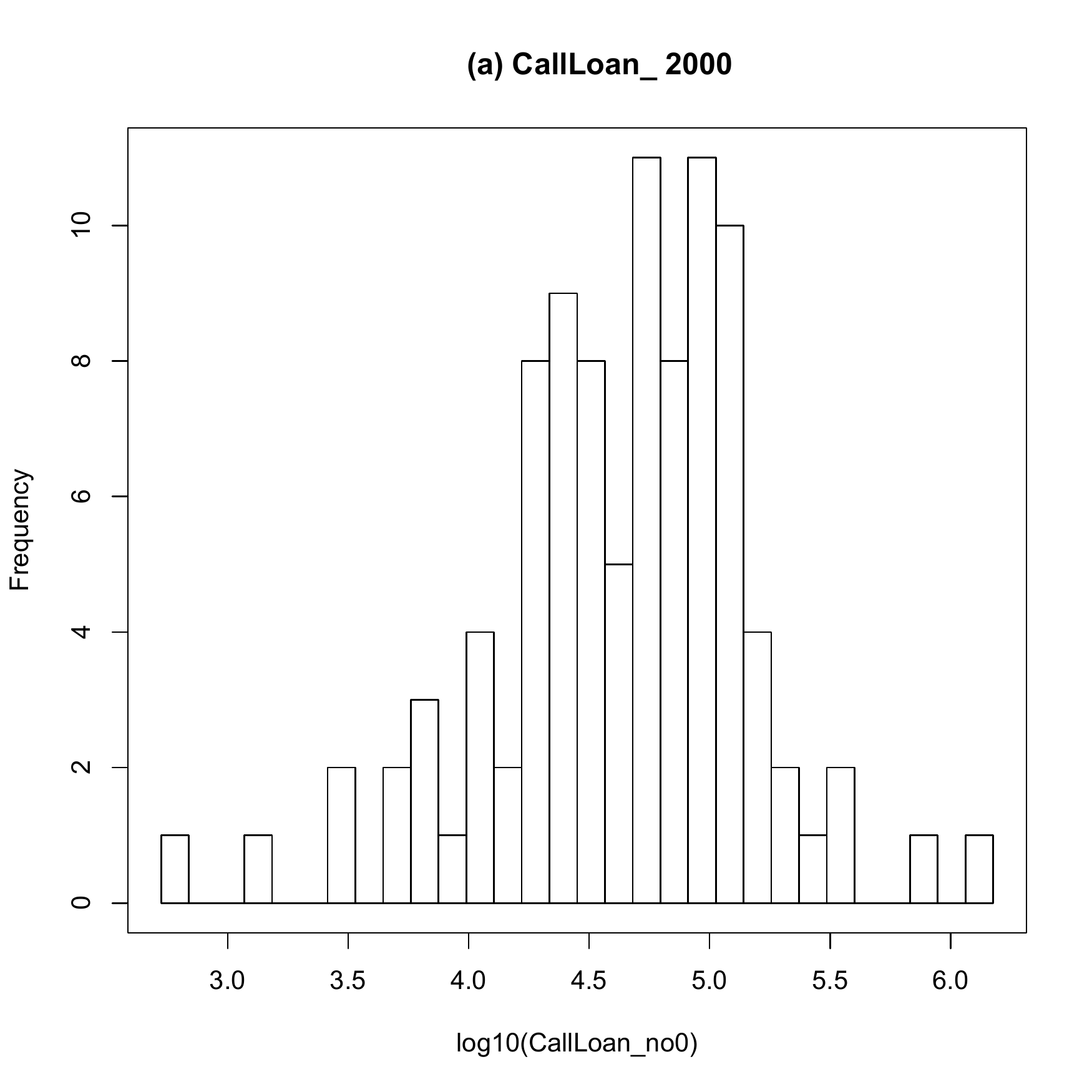}
  \includegraphics[width=0.4\textwidth]{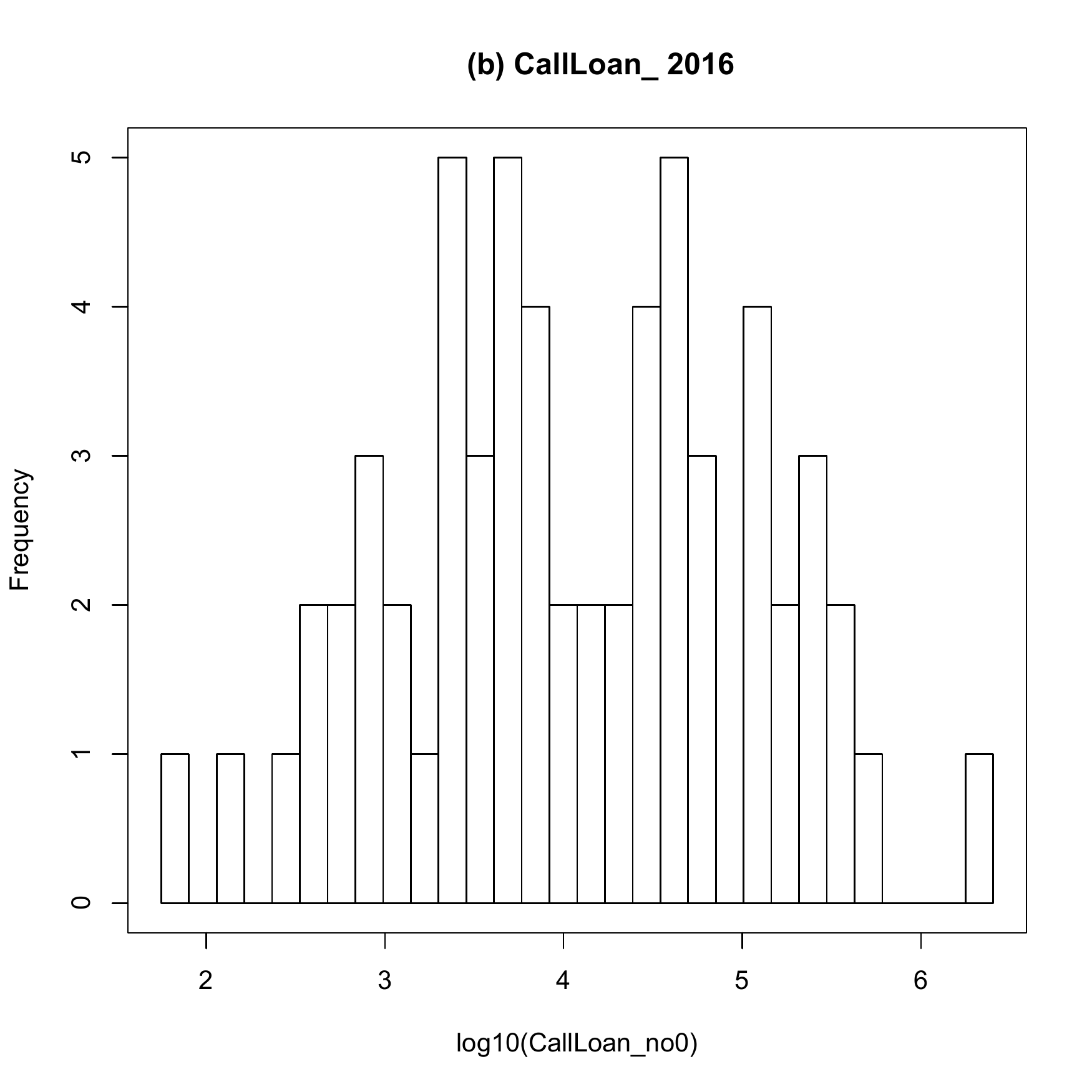}
\caption{Distribution of log-transformed call loan (lending) in 2000 (left panel) and 2016 (right panel). The aggregated call loan in balance sheet is consistent with the interbank market data, except for the transaction of the regional banks.}
\label{fig:LogCallLoan}       
\end{figure*}

\begin{figure*}
  \includegraphics[width=0.4\textwidth]{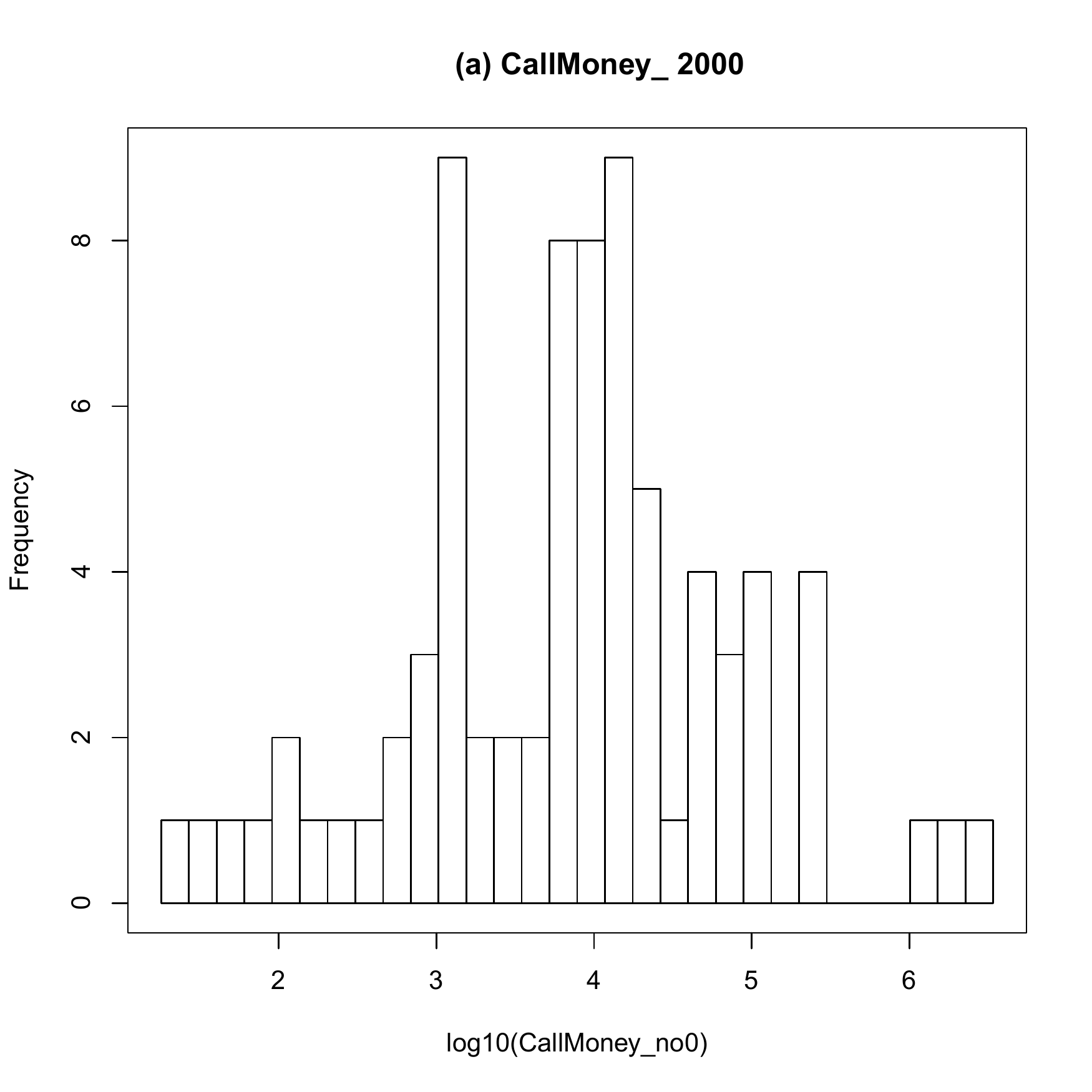}
  \includegraphics[width=0.4\textwidth]{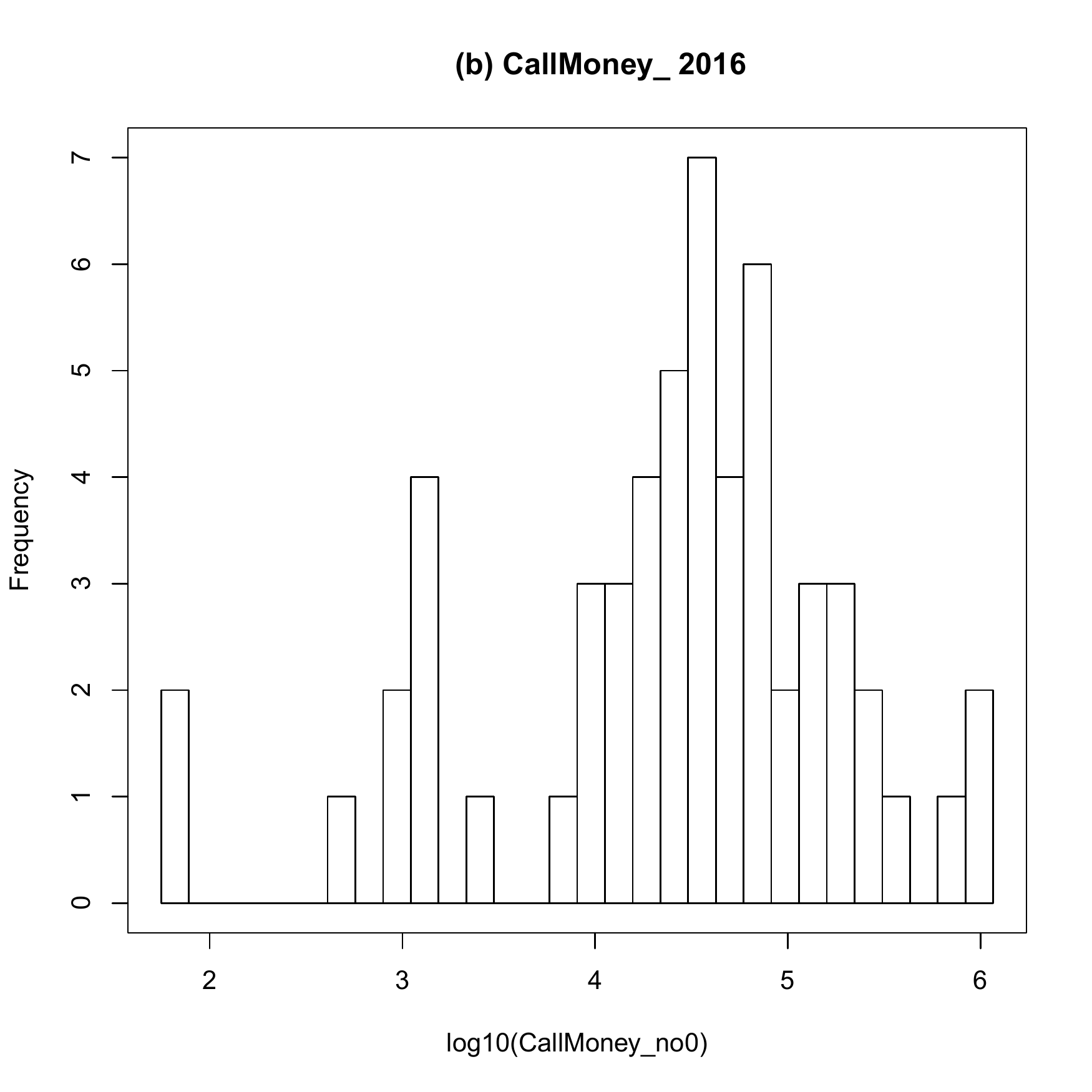}
\caption{Distribution of log-transformed call money (borrowing) in 2000 (left panel) and 2016 (right panel). The aggregated call money in balance sheet is consistent with the interbank market data, except for the transaction of the major banks.}
\label{fig:LogCallMoney}       
\end{figure*}

The distribution of log-transformed call loans in 2000 and 2016 are shown in Fig. \ref{fig:LogCallLoan}.
The distributions are regarded as a unimodal distribution approximately.
The distribution of log-transformed call money in 2000 and 2016 are shown in Fig.\ref{fig:LogCallMoney}.
The distributions are considered as a unimodal distribution approximately.
The amount of aggregate call loan is not equal to the amount of aggregate call money. Therefore, node ``other'' is needed as a slack variable in the ridge entropy maximization model formulated in Eqs. (\ref{ConvexFormulation}).

\paragraph{Comparison of Interbank Market and Balance Sheet}

Aggregated call loan (lending) and call money (borrowing) of the major banks, the trust banks, the regional banks, and the 2nd-tier regional banks recorded in balance sheets in 2005 are shown in Table \ref{table:BSCallLoanMoney}.
The aggregated call loan (lending) and the aggregated call money (borrowing) in the balance sheet are much smaller than the interbank market data.

Since the call money and call loans on the balance sheet are the balances at the end of the year of the transaction, it can be understood as natural that they are much smaller than the amount of call market transactions made during the month of December 2005.
The majority of transactions in the call market are overnight transactions, i.e., borrowing (or lending) on the same day and returning (or receiving repayment) the next day. It is quite natural that the amount of transactions expands daily, but the outstanding balance is considerably smaller than the amount of transactions.

If the amount of transactions in the call market is large, the amount of call loans on the balance sheet can be considered significant.
However, the picture changes when a financial crisis occurs or a policy change occurs in the period.
For example, in FY2008 (the subprime mortgage crisis in September 2008), FY2010 (the Bank of Japan supplied a large amount of funds to the market in response to the Great East Japan Earthquake in March 2011), and FY2015 (the introduction of negative interest rates in February 2016), there may be a discrepancy between the characteristics of transactions during the entire period and the ending balance.
In the absence of such special events, it is reasonable to say that the ending balance broadly reflects the characteristics of transactions during the period.

Therefore, in this study, we assume that the call money and call loans in the end-of-period balance sheet retain the characteristics of call market transactions.

\begin{table}[hbtp]
  \caption{Balance Sheet in 2005 ($10^{12}$ JPY) }
  \label{table:BSCallLoanMoney}
  \centering
  \begin{tabular}{lcc}
    \hline
    Category  &  Call Loan  & Call Money \\
    \hline \hline
    Major bank        & 1.5 & 9.4 \\
    Trust bank        & 0.5 & 0.6 \\
    Leading regional & 2.7 & 1.5 \\
    2nd tier regional & 0.4 & 0.1 \\
    \hline
  \end{tabular}
\end{table}

\section{Results and Discussions}
\label{sec:5}

The interbank network in Japan was reconstructed using the ridge entropy maximization model in Eqs. (\ref{ConvexFormulation}).
The number of banks in each category isis 5, 59, 3, and 31: the major commercial bank, the leading regional bank, the trust bank, and the second-tier regional bank, respectively. 
Call loan $s_{i}^{out}$ of bank $i$ and call money $s_{j}^{in}$ of bank $j$ are taken from the balance sheet of each bank and are given as constraints of the model.
In addition to the banks, a slack variable is incorporated in the model to balance the aggregated call loan and the aggregated call money. 
The sparsity that is no transactions within the same bank category except for major commercial banks, observed in Table \ref{table:CallLoanMoney} is considered by imposing the constraints in Eq. (\ref{GroupC2}), i.e.  $p_{ij} =0  \left( i \in g_l, j \in g_l \right)$, where $l$ denote the trust banks, the regional banks, and the 2nd-tier regional banks.
In the objective function in Eqs. (\ref{ConvexFormulation}), we assumed $\beta = 100$.
The optimization was done using package software for convex programming problem CVXR in R \cite{CVXR}.

For the verification of the ridge entropy maximization model in Eqs. (\ref{ConvexFormulation}) by applying to the problem where the adjacent matrix and its weights are known, see Appendix A.

\paragraph{Reconstruction of Interbank Network}

\begin{figure*}
  \includegraphics[width=0.4\textwidth]{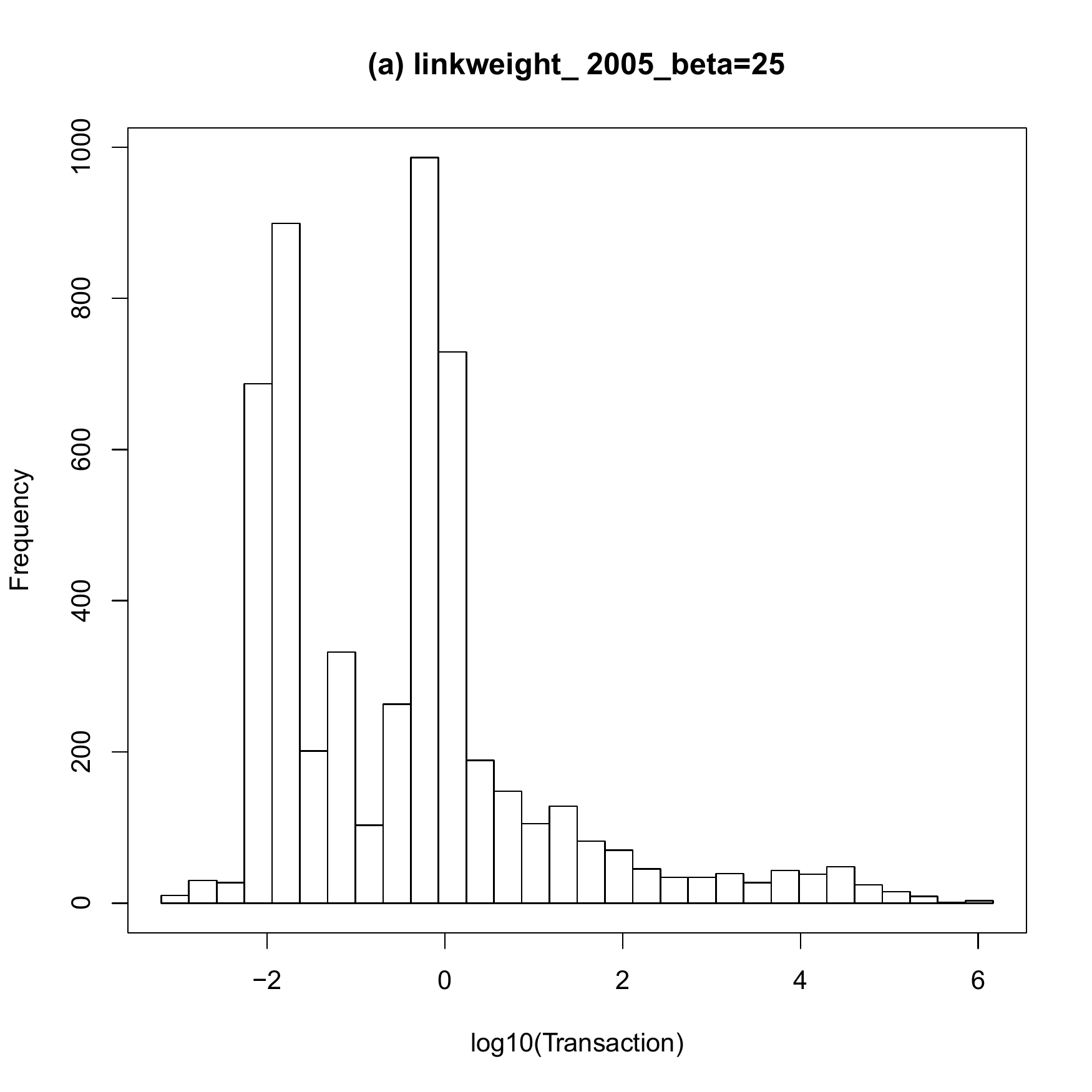}
  \includegraphics[width=0.4\textwidth]{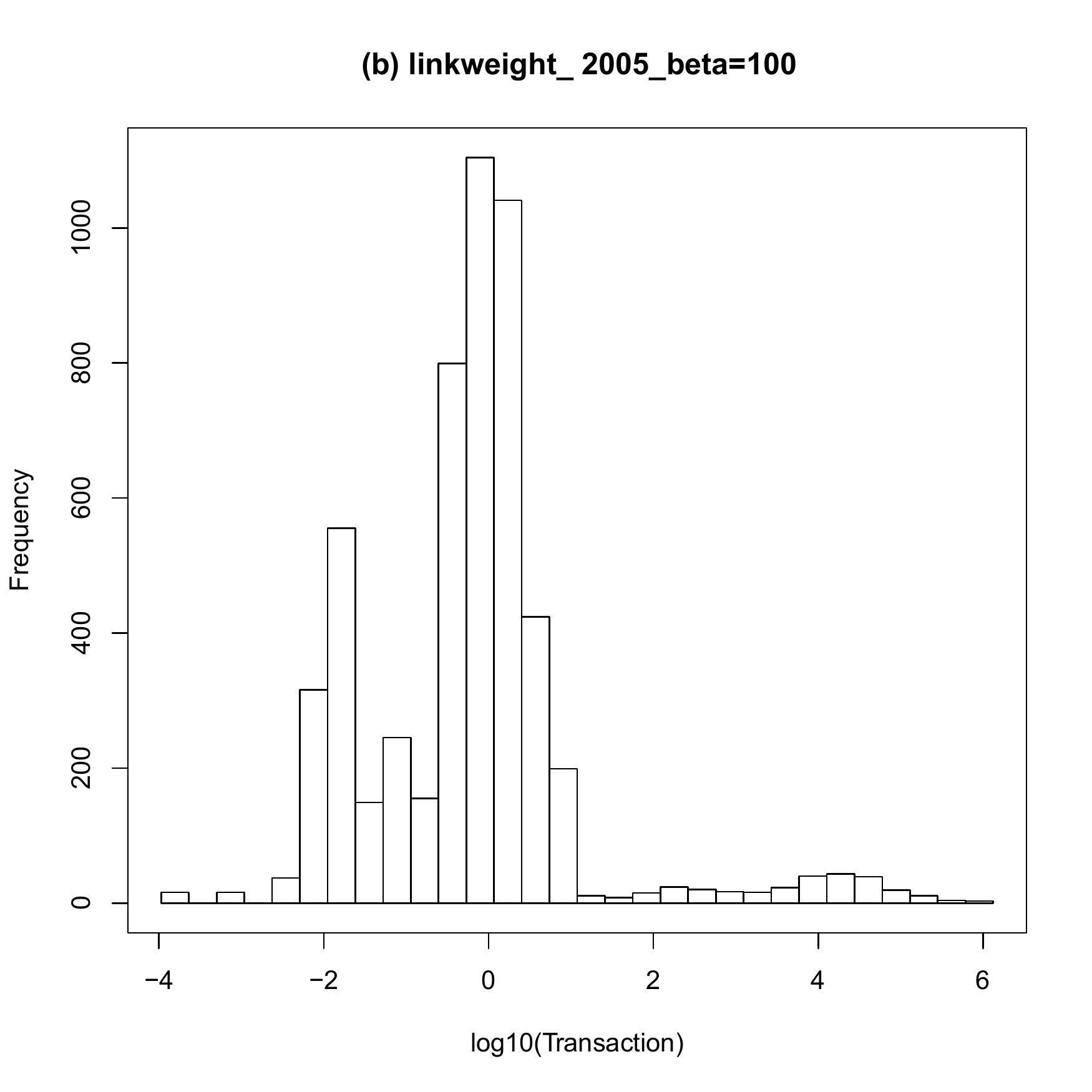}
\caption{Reconstructed Interbank Network in 2005. The left panel shows the distribution of weight $t_{ij}$ for the obtained interbank network with $\beta=25$. The right panel shows the distribution of weight $t_{ij}$ with $\beta=100$. 
Increasing $\beta$ resulted in a decrease in the links for small values and an increase in the large region.}
\label{fig:BankNet}       
\end{figure*}

The distributions of transaction $t_{ij}$ for the reconstructed interbank network in 2005 with the different value of $\beta$ are shown in Fig. \ref{fig:BankNet}. 
The leftmost peaks in the distributions are close to zero and therefore are regarded as spurious links.
The left panel shows the distribution of weight $t_{ij}$ for the obtained interbank network with $\beta=25$. The right panel shows the distribution of weight $t_{ij}$ with $\beta=100$. 
The comparison confirmed that increasing $\beta$ resulted in a decrease in the links for small values and an increase in the large region.

\paragraph{Binary Interbank Network by truncating Links}

\begin{figure*}
  \includegraphics[width=0.45\textwidth]{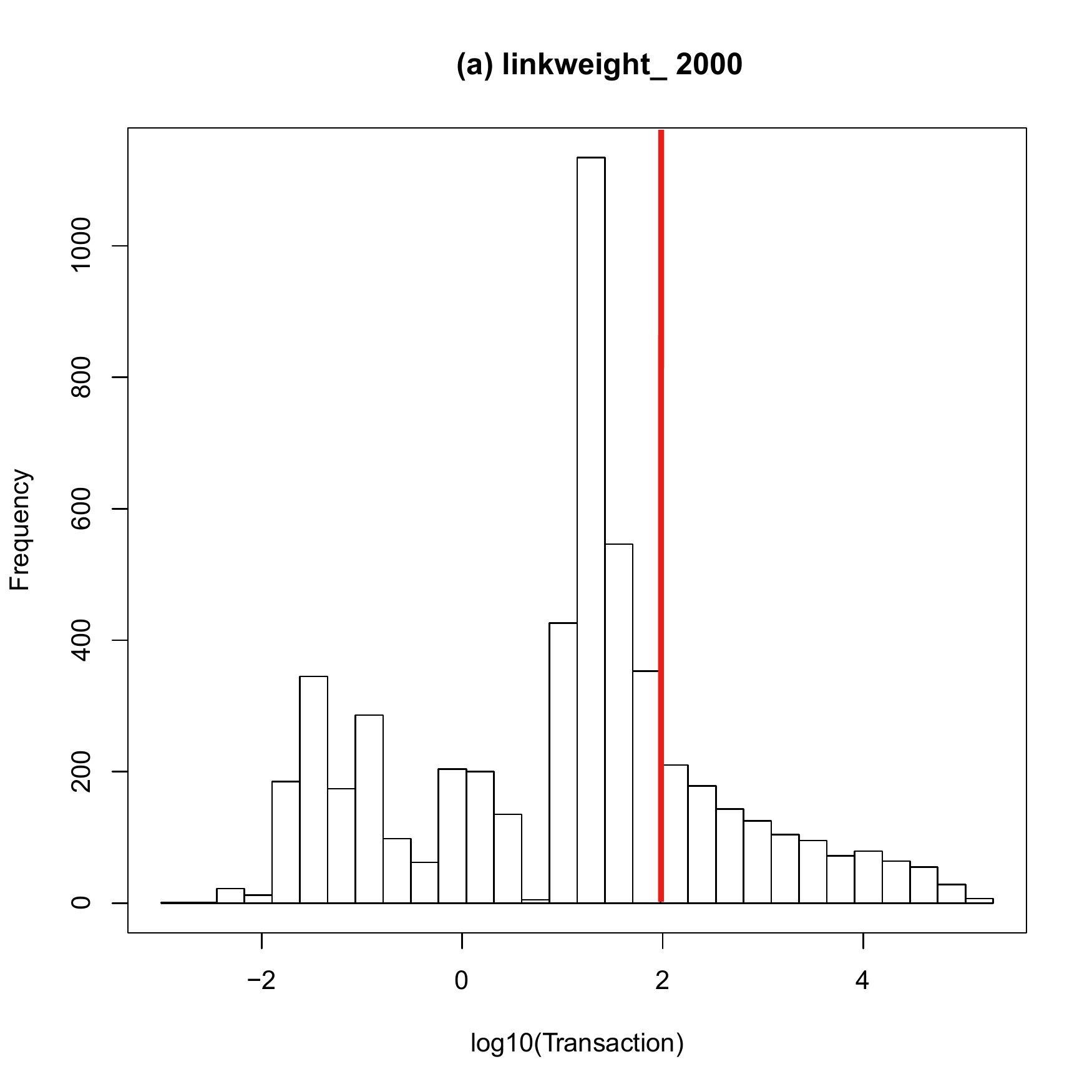}
  \includegraphics[width=0.45\textwidth]{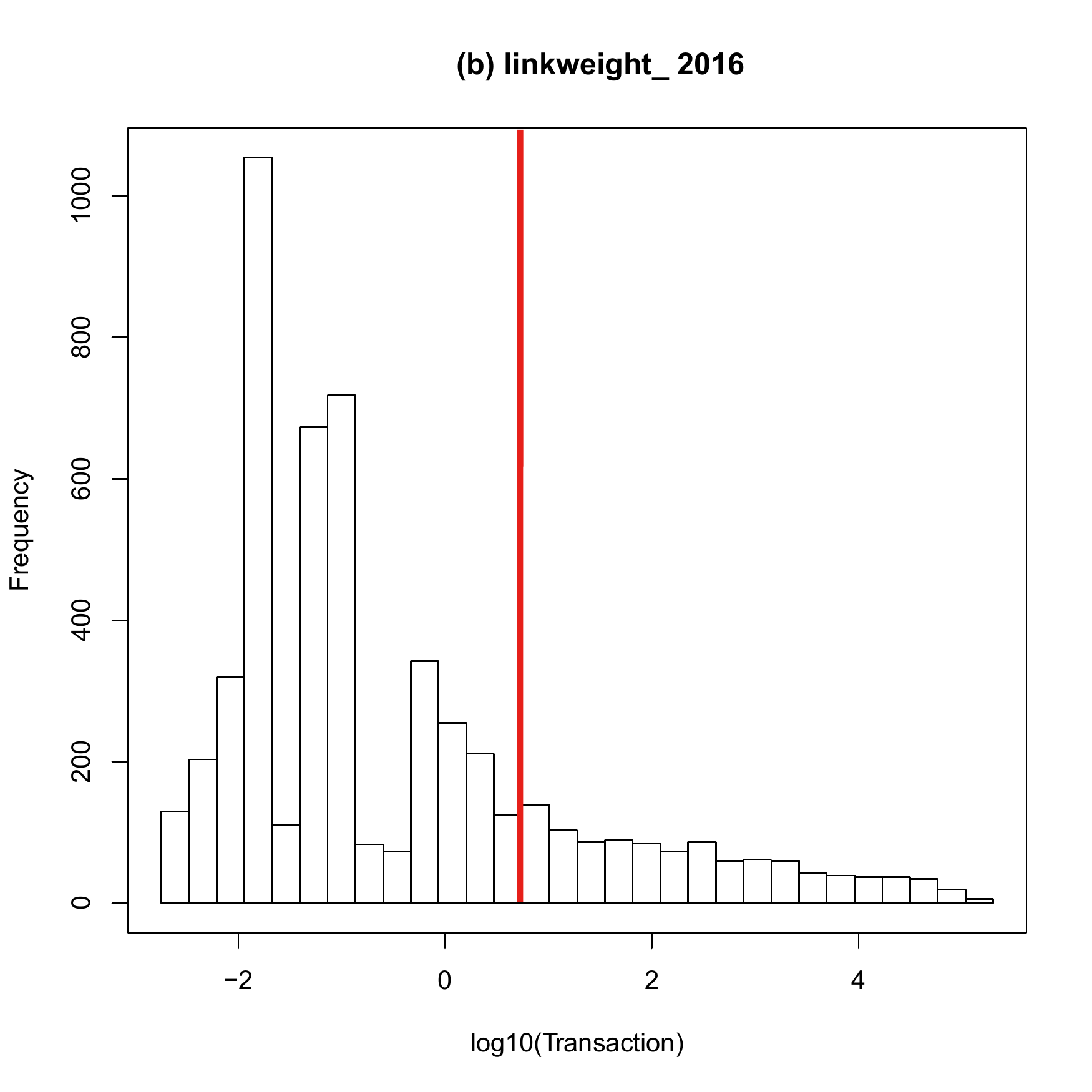}
\caption{The distributions of reconstructed transaction $t_{ij}$ are shown for 2000 in the left panel and for 2016 in the right panel. The red vertical line shows the $80^{th}$ percentile of the distribution and therefore the links smaller than this threshold are ignored in the binary interbank network.}
\label{fig:DistTrans}       
\end{figure*}

Binary interbank network $a_{ij}$ was obtained by truncated links whose weights $t_{ij}$ are smaller than the $80^{th}$ percentile of the distribution:
\begin{equation}
a_{ij} = 
\begin{cases}
    1,  & t_{ij} > \text{the } 80^{th} \text{ percentile} \\
    0,  & t_{ij} \le \text{the } 80^{th} \text{ percentile}.
\end{cases}
\label{Truncation}
\end{equation}
The distributions of reconstructed transaction $t_{ij}$ are shown in Fig. \ref{fig:DistTrans}.
The left panel is the distribution in 2000 and the right panel in 2016.
The red vertical line shows the $80^{th}$ percentile of the distribution and therefore the links smaller than this threshold are ignored in the binary interbank network.
By truncating the links below the threshold, the network density decreased from $54.1\%$ to $10.7\%$; this network density decrease is stable over the period 2000-2016.

\paragraph{Network Attributes and Community Structure}

\begin{figure*}
  \includegraphics[width=0.45\textwidth]{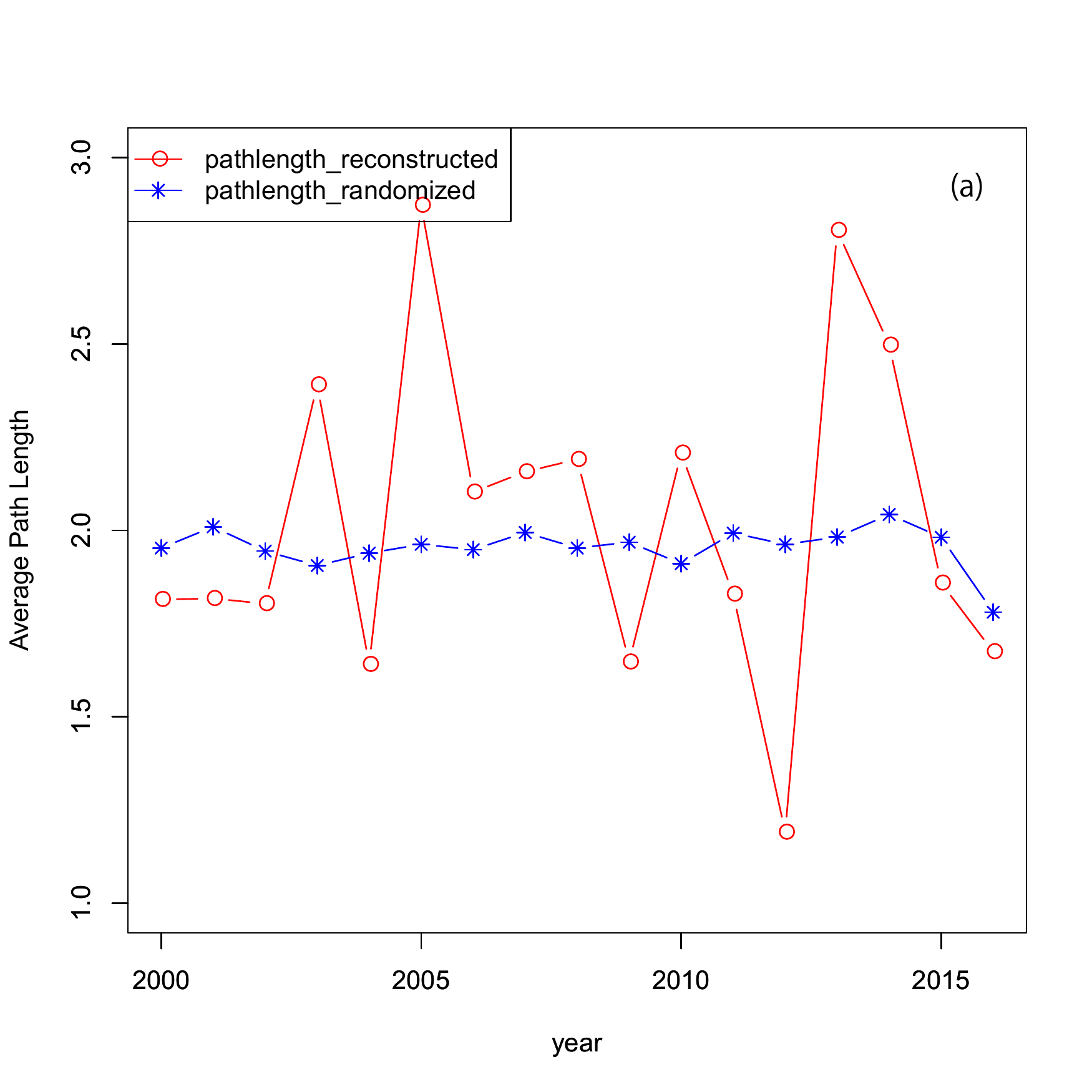}
  \includegraphics[width=0.45\textwidth]{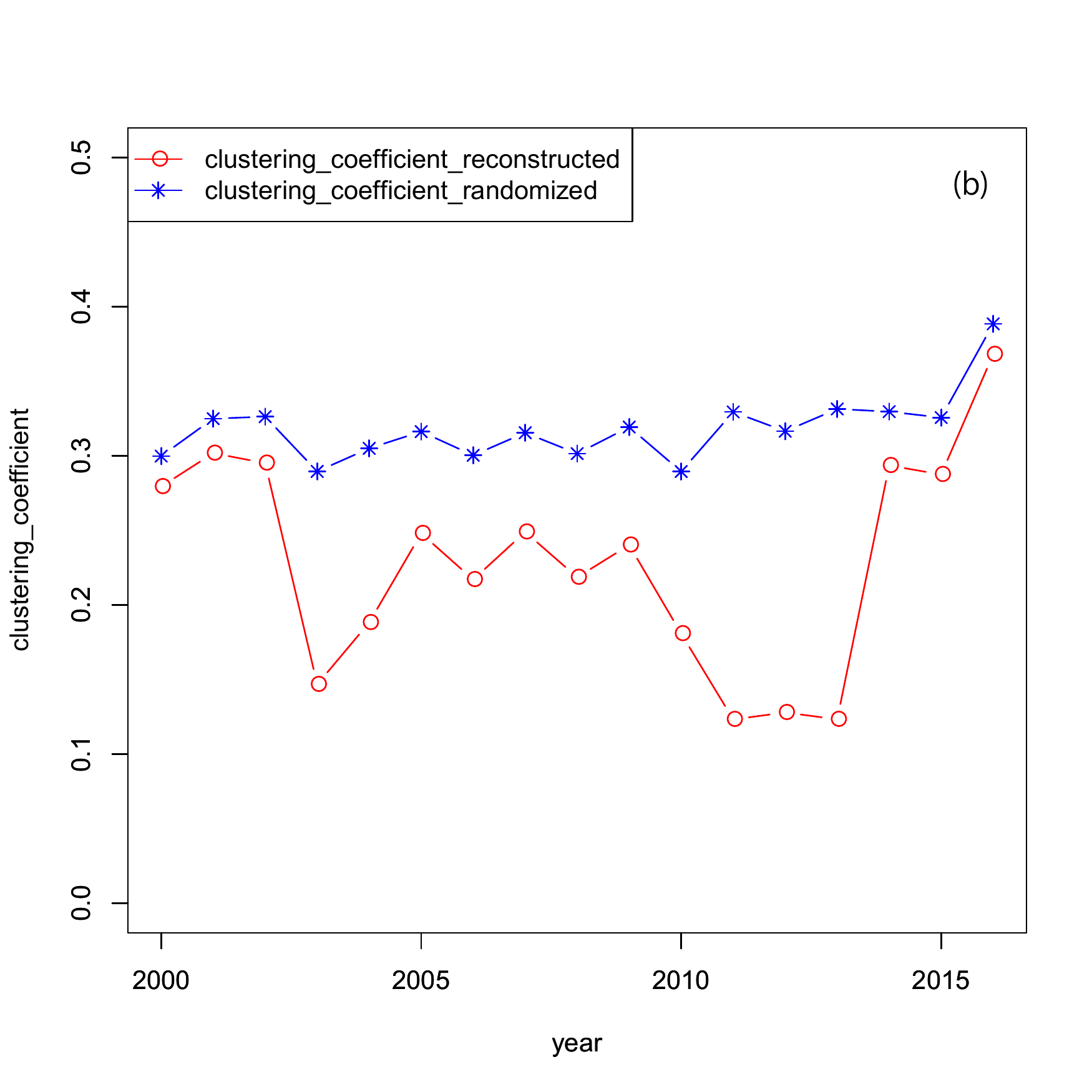}

  \includegraphics[width=0.45\textwidth]{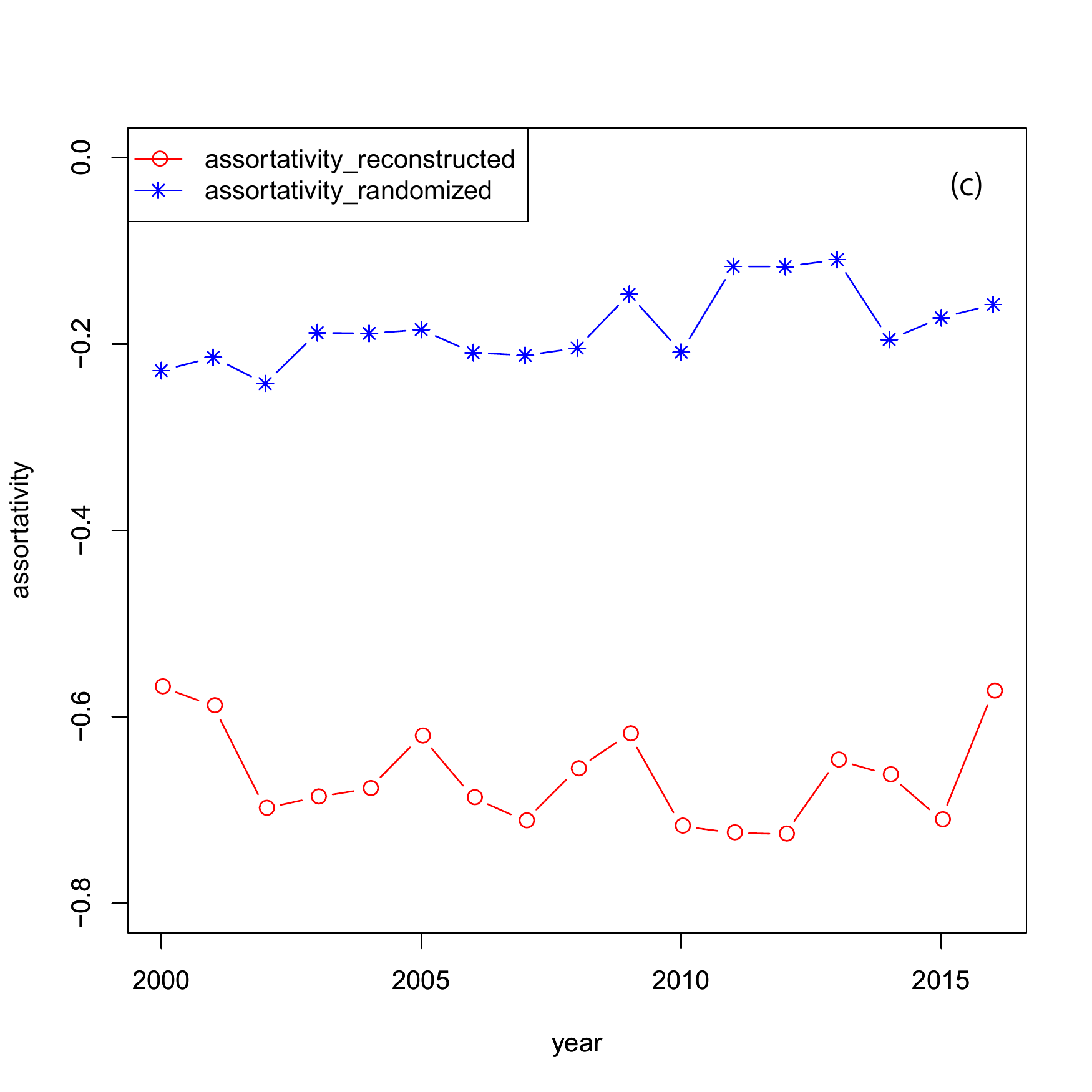}
  \includegraphics[width=0.45\textwidth]{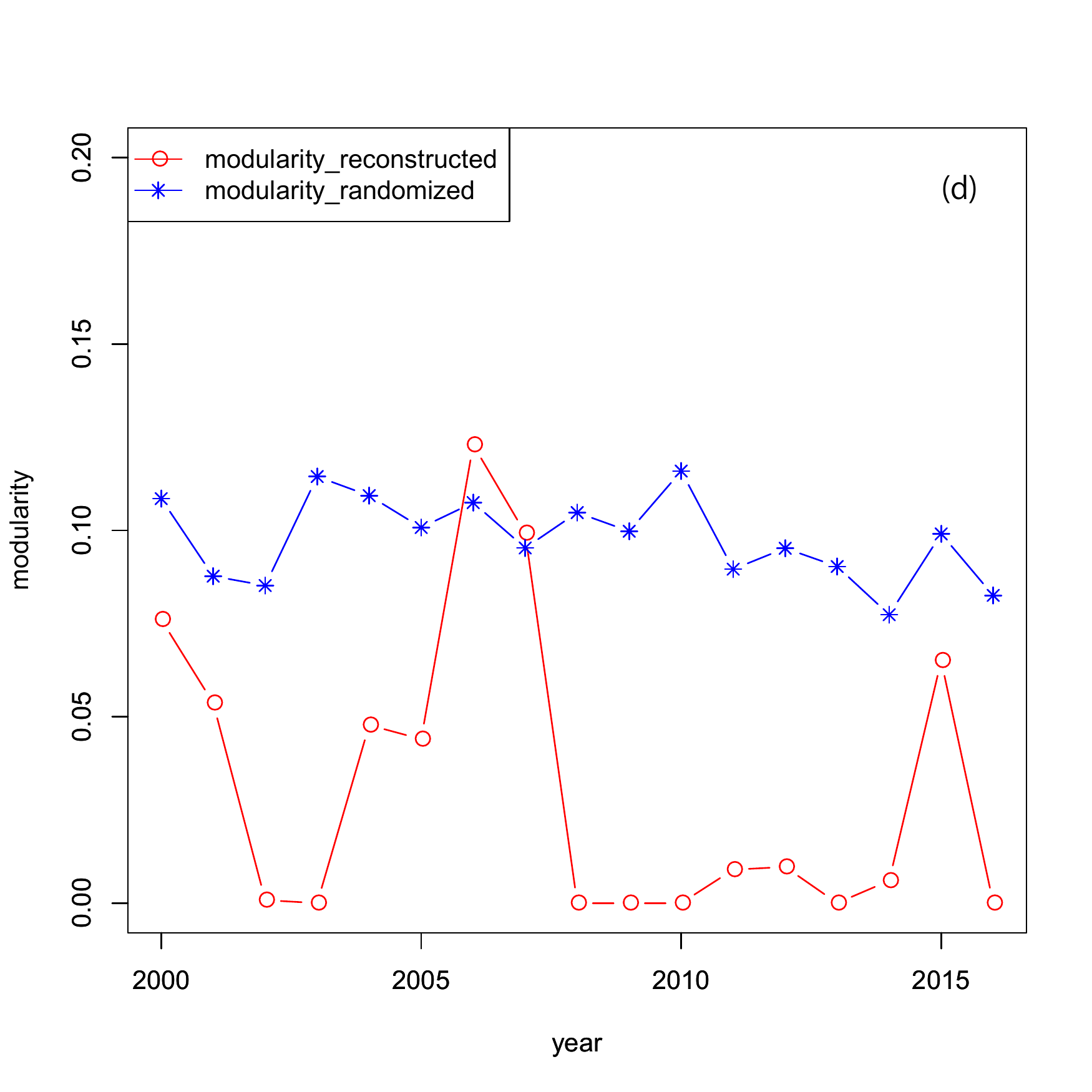}
\caption{The temporal change of the network attributes: 
The average shortest path length, the clustering coefficient, the assortativity, and modularity in degree-preserved randomized network are shown for the binary network reconstructed from 2000 to 2016.
The results comfirmed that the reconstructed network is essentially random network.
The comparison of the assortativity implies that the strong disassortativity emerged in the reconstructed interbank network. }
\label{fig:RandNet}       
\end{figure*}

Three centralities: the clustering coefficient, the path length, and the assortativity, were calculated for the reconstructed binary network from 2000 to 2016.
The community structures were identified by maximizing the modularity for the binary networks.
The temporal change of the network attributes: the modularity, the clustering coefficient (transitivity), the assortativity, and the shortest path length are shown in panels (a)-(c) of Fig. \ref{fig:RandNet} between 2000 and 2016.

The observed average shortest path in panel (a) of Fig. \ref{fig:RandNet} is approximately $2$, which is consistent with the previous study \cite{Muller2006}.
Panel (b) and (c) of Fig. \ref{fig:RandNet} also shows a small clustering coefficient and disassortative property, respectively.
The strong disassortativity implies that a few large banks are linked to a large number of small banks.
These are also consistent with the previous studies \cite{Iori2008}, and \cite{Soramaki2006}.
Therefore, the reconstructed interbank network reproduced the stylized facts known for the interbank networks.
Panel (d) of Fig. \ref{fig:RandNet} shows that the modularity is less than $0.1$ during the entire period. This means that the identified communities are not necessarily the most significant, but they can be helpful to understand the network structure.

1000 samples of degree-preserved randomized networks were generated and
the clustering coefficient, the shortest path length, and the assortativity were calculated for the randomized networks.
Comparison with the original reconstructed network for the clustering coefficient, the shortest path length, and the assortativity are shown in panels (a)-(c) of Fig. \ref{fig:RandNet}.
The shortest path length and the clustering coefficient are equivalent during the entire period.
The disassortativity becomes weaker in the randomized network during the entire period.
The comparison of the disassortativity means that the strong disassortativity emerged in the reconstructed interbank network.
The rest properties are similar to the random network as expected by the basic concept of the reconstruction model.

\paragraph{Core and Peripheral Structure}

\begin{figure*}
  \includegraphics[width=0.5\textwidth]{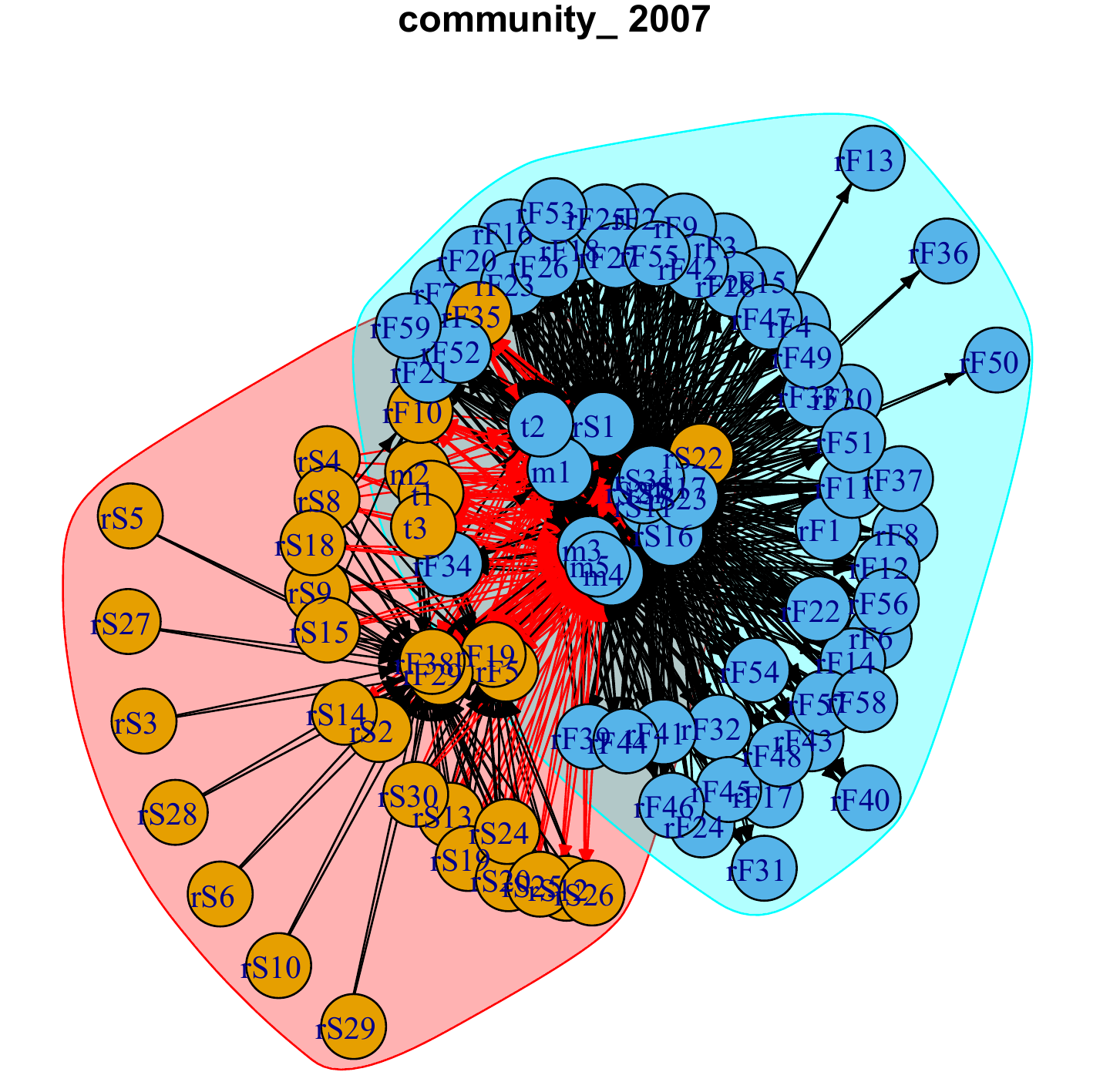}
  \includegraphics[width=0.5\textwidth]{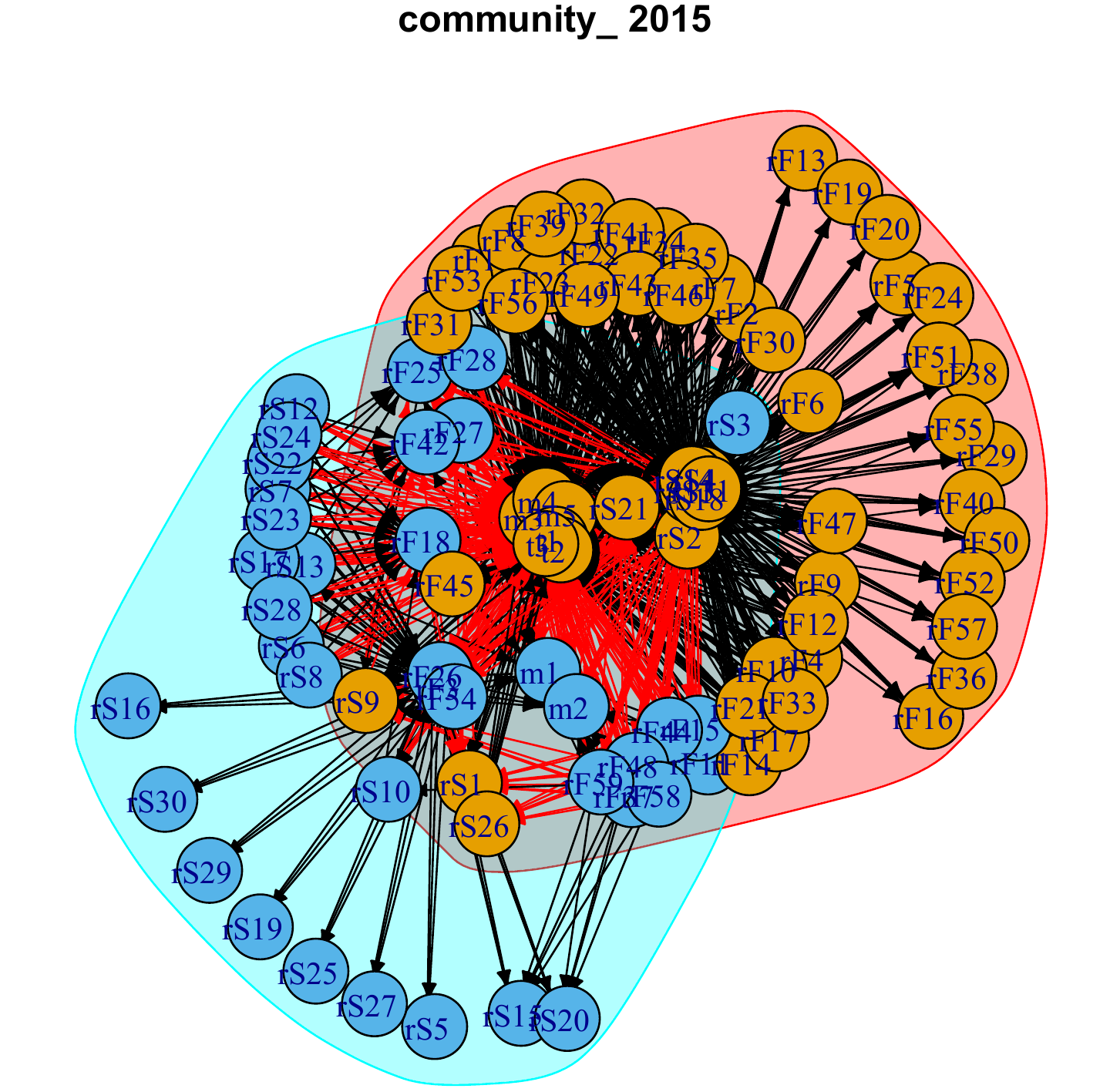}
\caption{Core and peripheral structure in 2007 (left) and 2015 (right). The area shaded by red or blue region forms major communities. In the core part, a few large banks  are interlinked each other and are linked to a large number of banks located in the peripheral region.}
\label{fig:CorePeripheral}       
\end{figure*}

The core and peripheral structures have been known in the interbank network \cite{Imakubo2010}.
The network structure is depicted in Fig. \ref {fig:CorePeripheral} for the binary interbank network in 2007 and 2015.
The area shaded by the red or blue region of Fig. \ref {fig:CorePeripheral} forms major communities.
We note that many banks are surrounding the core part.
In the core part, a few large banks are interlinked and linked to a large number of banks located in the peripheral region.
We note that we did not introduce the mechanism to generate the core and peripheral structure, but we imposed the constraints $p_{ij} =0 \left( i \in g_l, j \in g_l \right)$ to consider the sparsity that is no transactions within the same bank category except for major commercial banks.
Therefore, we can say that the core and peripheral structure is spontaneously emerged.

\begin{figure*}
  \includegraphics[width=0.45\textwidth]{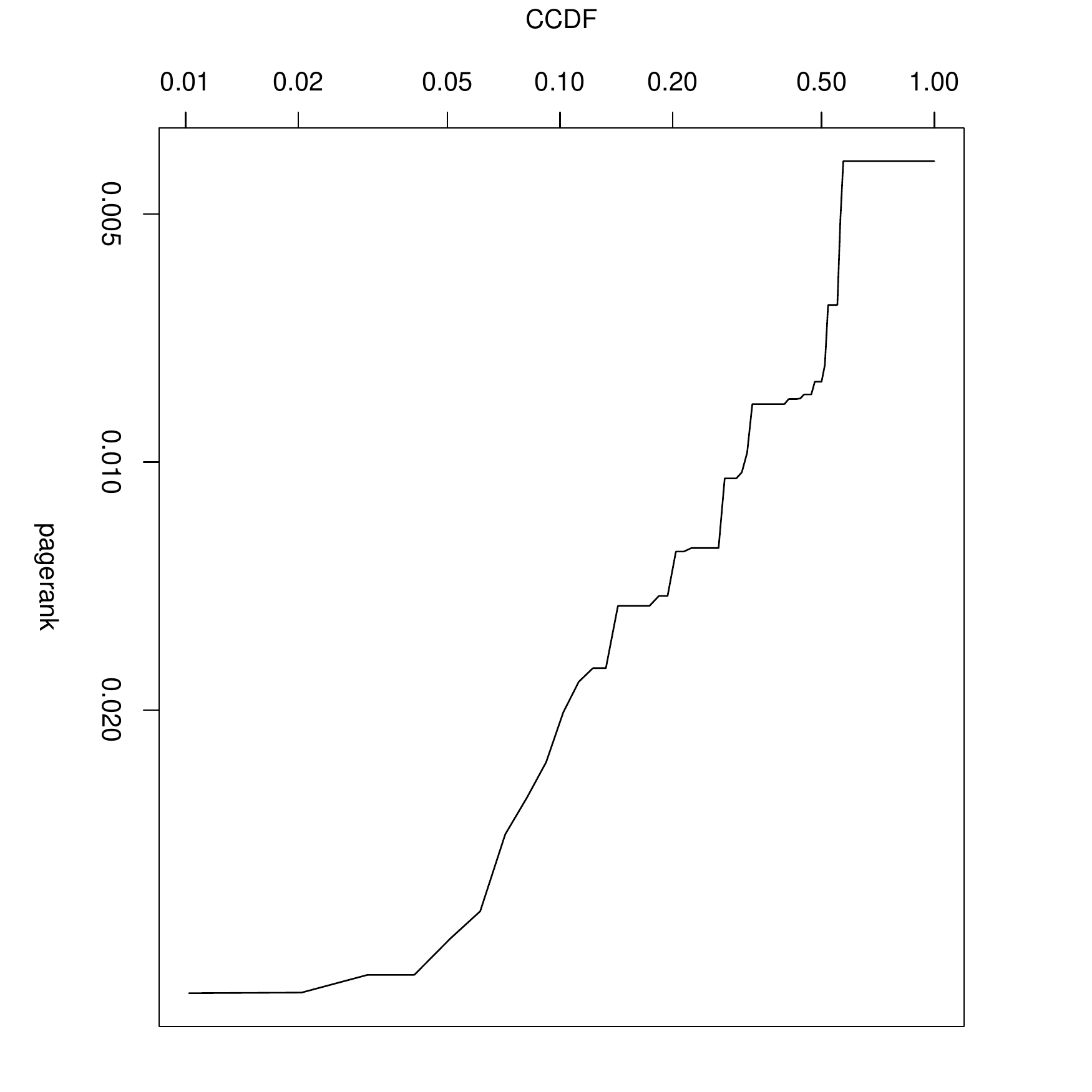}
  \includegraphics[width=0.45\textwidth]{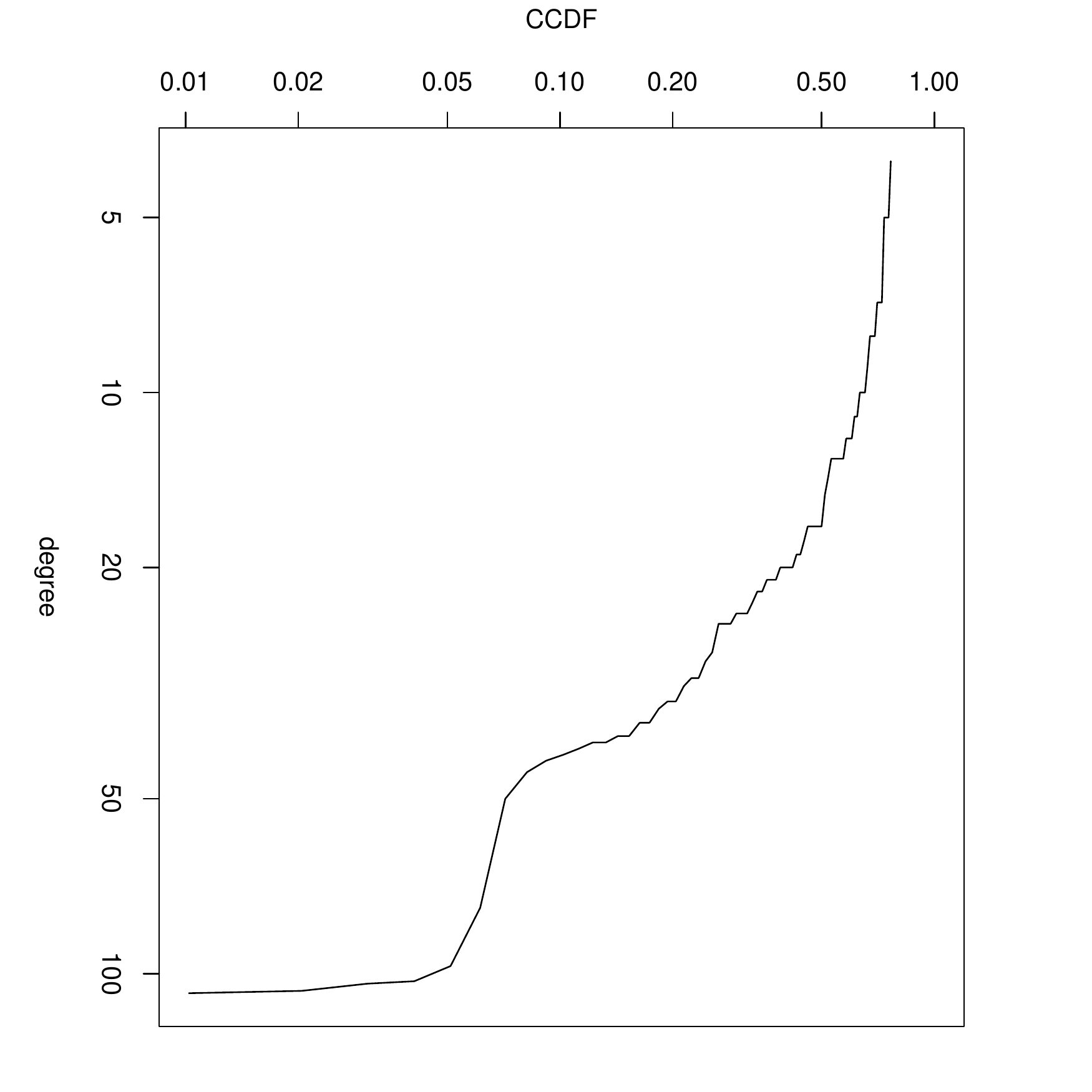}
\caption{Heterogeniety of nodes in PageRank distribution and degree distribution in 2016. The Complementary Cumulative Distribution Function (CCDF) imply that the heterogeniety of nodes is strong and thus major nodes could be identified using the value of PageRank and degree.}
\label{fig:Heterogeniety}       
\end{figure*}

Cumulative PageRank distribution and degree distribution in 2016 are shown in Figure \ref{fig:Heterogeniety}.
These distributions imply that the heterogeneity of nodes is substantial, and thus major nodes could be identified using the value of PageRank and degree.


\paragraph{Changing Role of Bank Categories}

Two large communities were detected in 2000, 2007, and 2015, which showed relatively large values of modularity. In these communities, we listed the ten financial institutions with the highest PageRank as major nodes. 
The major nodes of the first and second communities are shown in Table \ref{table:MN2000Comm1} to Table \ref{table:MN2015Comm2} in Appendix B.

From 2000 to 2015, Mizuho Bank, MUFG Bank, Aozora Bank, Snisei Bank, and Resona Bank belong to the major commercial bank category and are listed as major nodes. The number of in-degrees of these financial institutions is consistently much higher than the number of out-degrees, indicating that they are continuously raising funds through the call market.

In the major node in 2000, the in-degree of Shizuoka Bank, Bank of Yokohama, Joyo Ban, Kagoshima Bank, Gunma Bank, which belong to the leading regional bank category, is higher than the out-degrees.
In the major node of 2007, the number of in-degrees of Shimizu Bank, Nanto Bank, Daishi Bank, and Shonai Bank, which belong to the leading regional bank category, is higher than the out-degrees.
However, these banks belonging to the leading regional bank category are not included in the major node in 2015, or even if they are included, the number of in-degree and out-degree are often at the same level. These results indicate a change in the position of the leading regional bank category.

\begin{figure*}
  \includegraphics[width=0.45\textwidth]{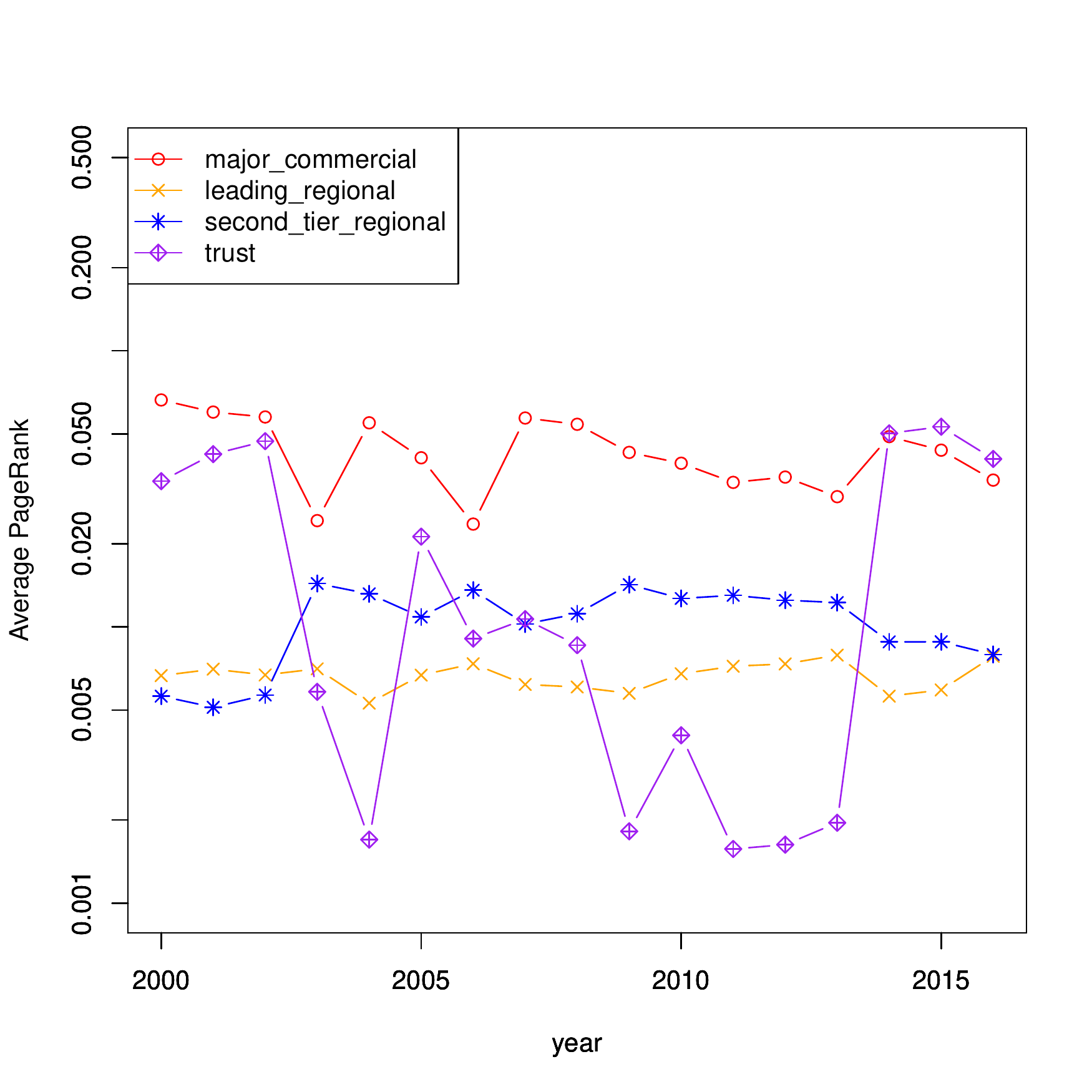}
  \includegraphics[width=0.45\textwidth]{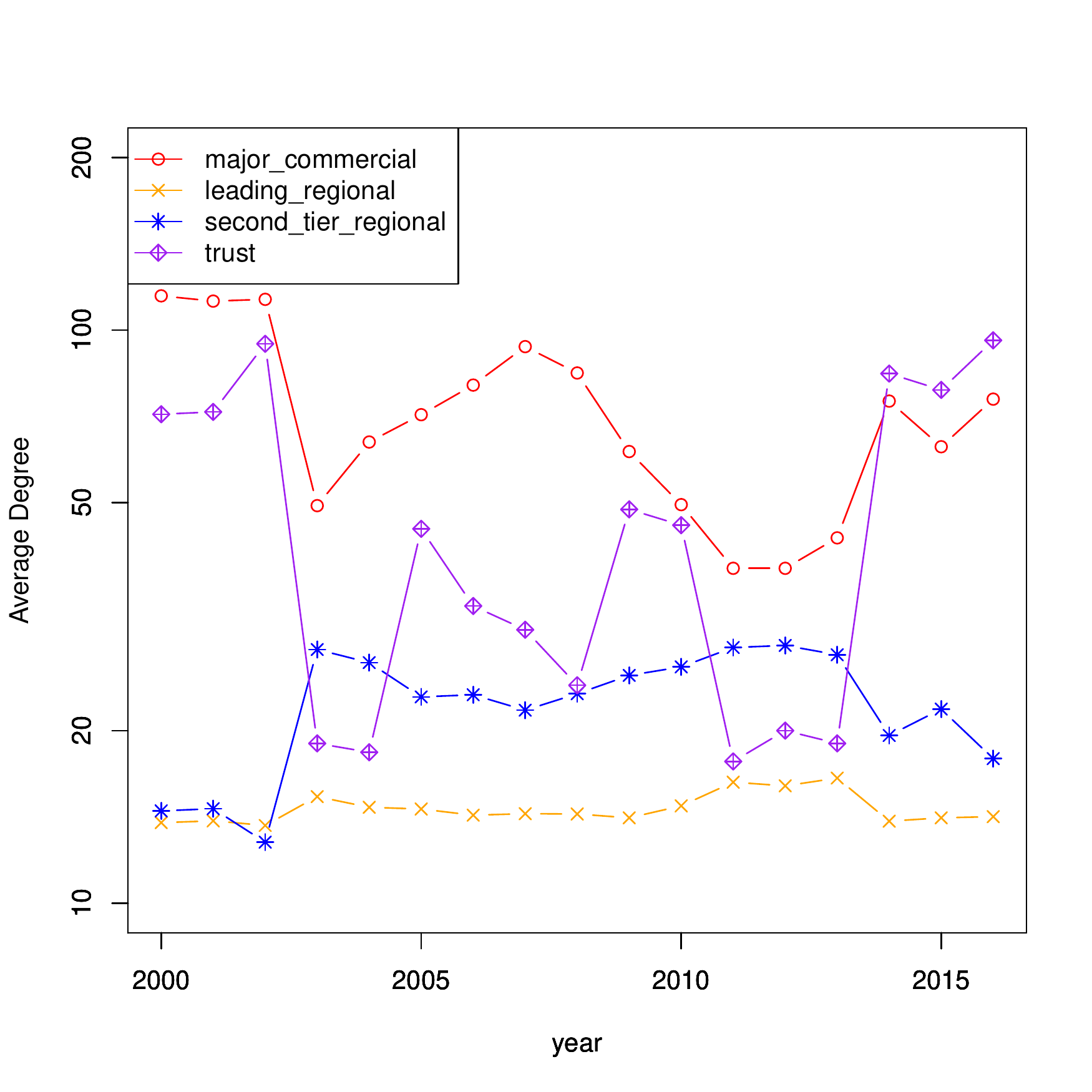}
  
  \includegraphics[width=0.45\textwidth]{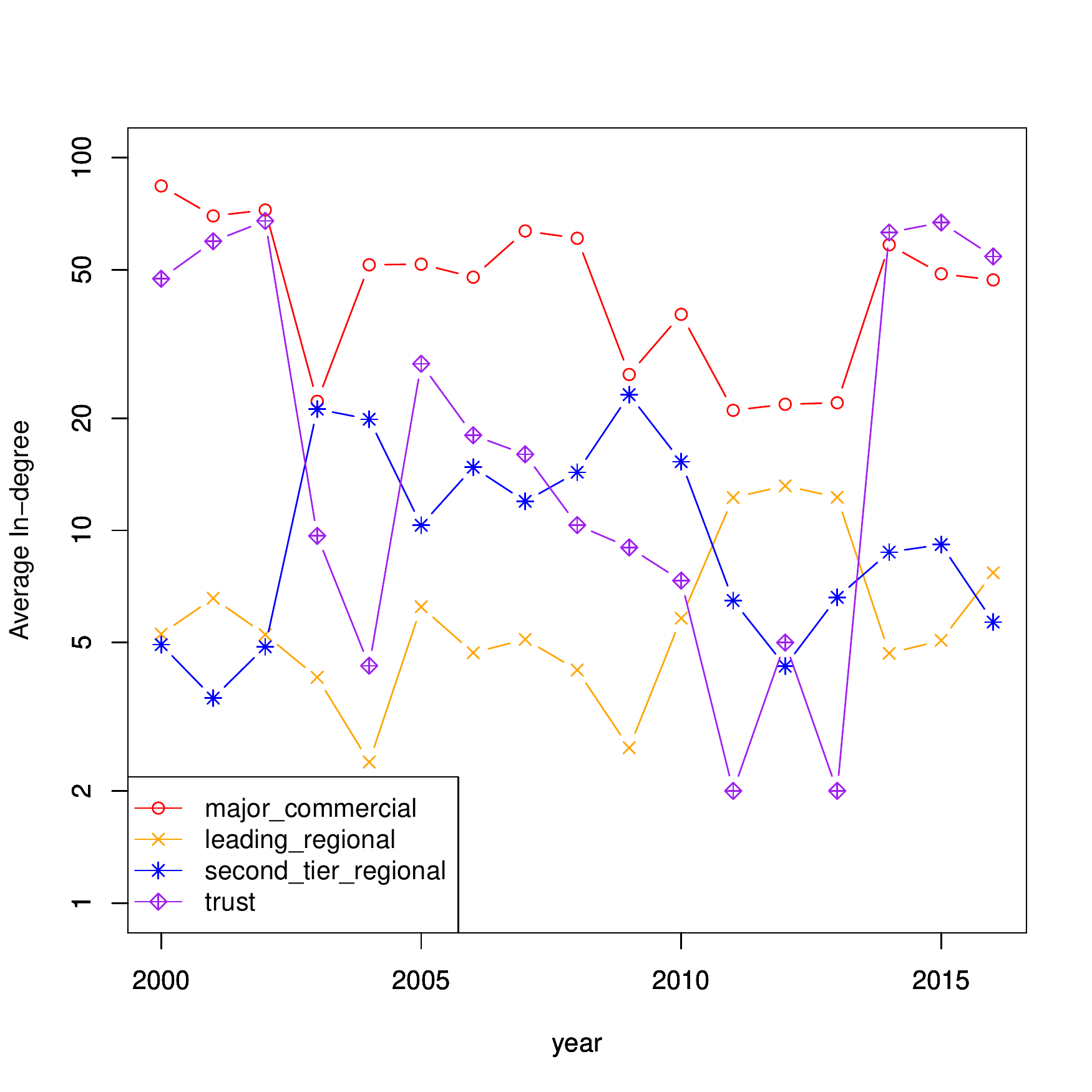}
  \includegraphics[width=0.45\textwidth]{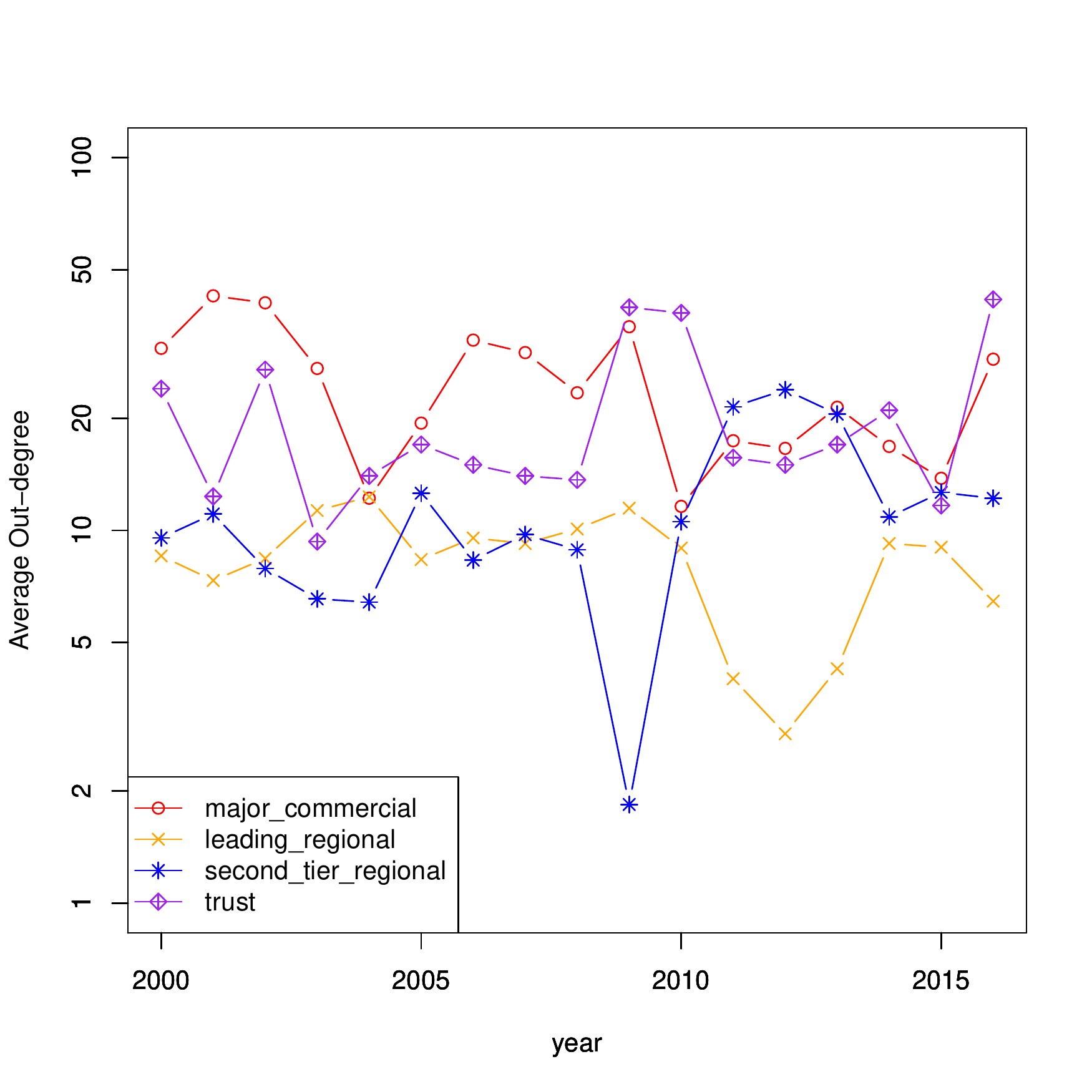}
\caption{Changing role of banks observed in PageRank, degree, in-degree, and out-degree. }
\label{fig:ChangingRole}       
\end{figure*}

The average PageRank, average degrees in each bank category are summarized in Fig. \ref{fig:ChangingRole}.
Since 2014, both PageRank and the degree of trust banks have increased significantly; the increase in degree may be mainly due to the increase in in-degree.
In addition, the role of second-tier regional banks in providing funds appears to have increased in recent years.
The observed changing role of banks is considered as a result of the quantitative and qualitative monetary easing policy started by the Bank of Japan in April of 2013.
Although a more detailed study is needed, it can be said that the ridge entropy maximization model in Eqs. (\ref{ConvexFormulation}) captured the change of the interbank network caused by the quantitative and qualitative monetary easing policy.

\section{Conclusion}
\label{sec:6}

We developed a network reconstruction model based on entropy maximization considering the sparsity of networks.
Here the reconstruction is to estimate the network's adjacency matrix from the node's local information.
We reconstructed the interbank network in Japan from financial data in balance sheets of individual banks using the developed reconstruction model in the period from 2000 to 2016.
The observed sparsity of the interbank network was successfully reproduced.

We examined the characteristics of the reconstructed interbank network by calculating important network attributes.
We obtained the following characteristics, which are consistent with the previously known stylized facts:
the short path length \cite{Muller2006}, the small clustering coefficient \cite{Iori2008}, the disassortative property \cite{Soramaki2006}, and the core and peripheral structure \cite{Imakubo2010}.
Although we did not introduce the mechanism to generate the core and peripheral structure, we imposed the constraints to consider the sparsity that is no transactions within the same bank category except for major commercial banks, the core and peripheral structure has spontaneously emerged.

Community analysis showed that two large communities were detected in 2000, 2007, and 2015, which showed relatively large values of modularity.
Major nodes in each community were identified using the value of PageRank and degree.
This identified that the major commercial banks and the trust banks were the major nodes in each community.

Since 2014, both PageRank and the degree of trust banks have increased significantly; the increase in degree may be mainly due to the increase in in-degree. In addition, the role of second-tier regional banks in providing funds appears to have increased in recent years.
The observed changing role of banks is considered a result of the quantitative and qualitative monetary easing policy started by the Bank of Japan in April of 2013.

\section*{Appendix A}
\label{sec:7}

In appendix A, we verify the ridge entropy maximization model in Eqs. (\ref{ConvexFormulation}) by applying it to the problem where the adjacent matrix and its weights are known.
We aggregated World Input-Output Data \cite{Timmer2015} in 2014 into country-wise trade data, which are expressed in millions of US dollars.
The aggregated data include 44 countries.
We prepared two different data sets $t_{ij}^{(data)}$: one is the original aggregated country-wise trade data denoted by No-cut data, and the other is the aggregated country-wise trade data deleted below the second quantile point, denoted by 2Q-cut data. 
Each row corresponds to the amount of export from the country shown in the row to the countries shown in columns.
Therefore, row-wise sum and column-wise sum correspond to the country's aggregated export $E_i$ and import $I_j$, respectively. 
We reconstruct the trade network by treating $E_i$ and $I_j$ as $s_i^{out}$ and $s_j^{in}$, respectively. 

\begin{figure*}
  \includegraphics[width=1.0\textwidth]{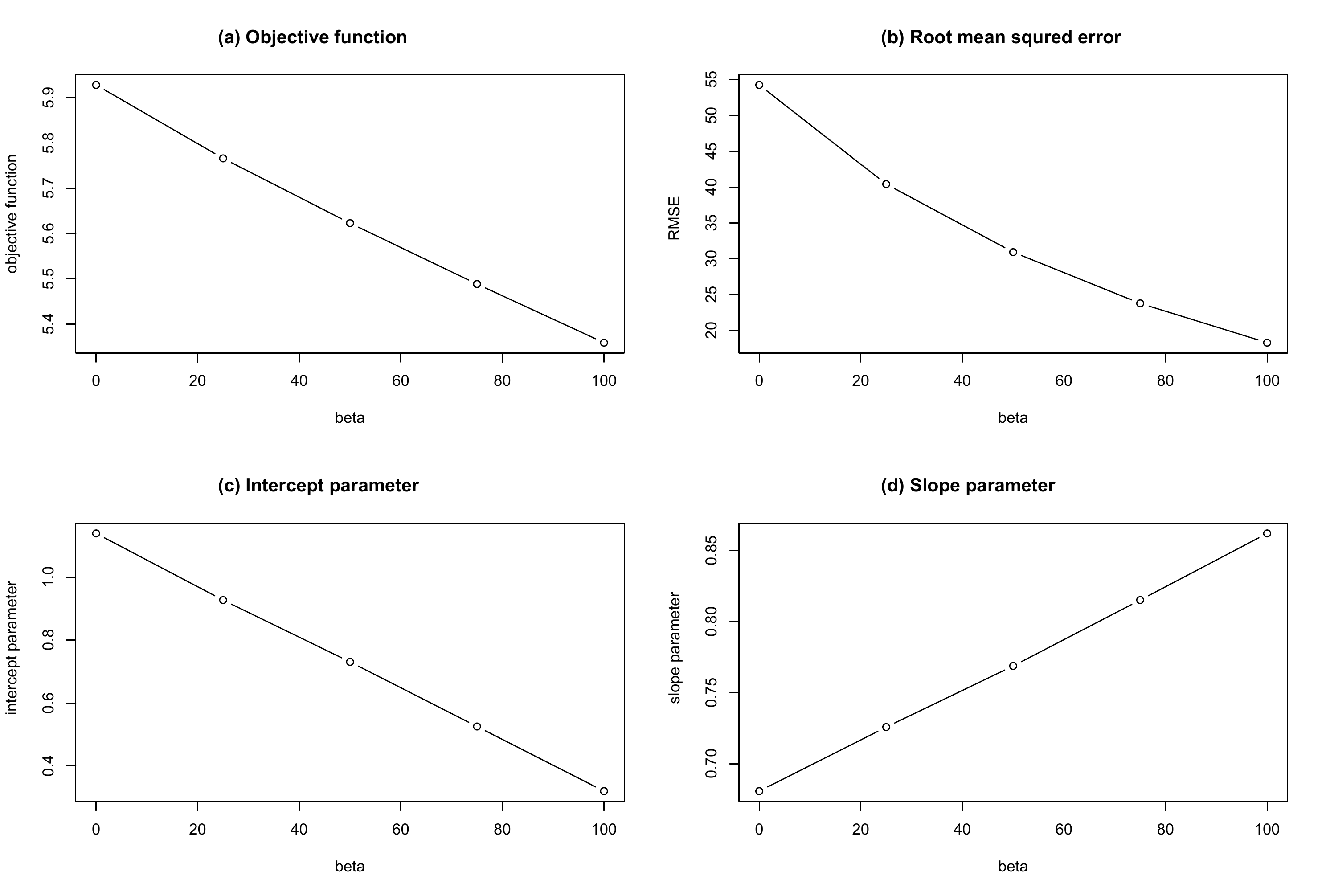}
\caption{The parameter $\beta$ dependences of the reconstruction for No-cut data without link constraints}
\label{fig:nocut_wo_lc}       
\end{figure*}

First, we reconstruct the trade network $t_{ij}^{(reconst)}$ for No-cut data without the link constraints. 
Here, the link constraints mean the additional constraints taken from the adjacent matrix and its weights in the Ridge-Entropy maximization.
We calculated the normalized root mean squared error between the trade data and the reconstructed trade,
\begin{equation}
  \mathrm{RMSE} = \sqrt{ \frac{1}{n} \sum_{ij} \left( \frac {t_{ij}^{(reconst)}-t_{ij}^{(data)}} {t_{ij}^{(data)} } \right)^2 }.
\label{RMSE}
\end{equation}
We also fitted the relationship between between the trade data and the reconstructed trade,
\begin{equation}
  \log_{10} { t_{ij}^{(reconst)} } = a \log_{10} t_{ij}^{(data)} + b.
\label{fitting}
\end{equation}
The parameter $\beta$ dependences of the reconstruction are shown in Fig. \ref{fig:nocut_wo_lc}.
Fig. \ref{fig:nocut_wo_lc} (a) to (d) show that the objective function $z(p_{ij})$ in Eq. (\ref{ConvexFormulation}), $\mathrm{RMSE}$, the intercept parameter $b$ of the fitting between the trade data and the reconstructed trade, and the slope parameter $a$ of the fitting, respectively.
As  $\beta$ is increased, $\mathrm{RMSE}$ decreases, the intercept $b$ approaches $0.0$, and the slope $a$ approaches $1.0$, although the objective function decreases.

Pure entropy maximization, without the second term of the objective function in Eq. (\ref{ConvexFormulation}), corresponds to $\beta=0$. In this case, under the given constraints, the link weights are relatively uniformly and randomly distributed. However, as $\beta$ is increased, the second term of the objective function increases the heterogeneity of the link weights. As a result, the slope parameter $a$ becomes larger; if the value of $\beta$ is too large, the slope parameter $a$ takes a value greater than one. Therefore, there is an optimal value of $\beta$ that best reproduces the heterogeneity of the link weights.

\begin{figure*}
  \includegraphics[width=1.0\textwidth]{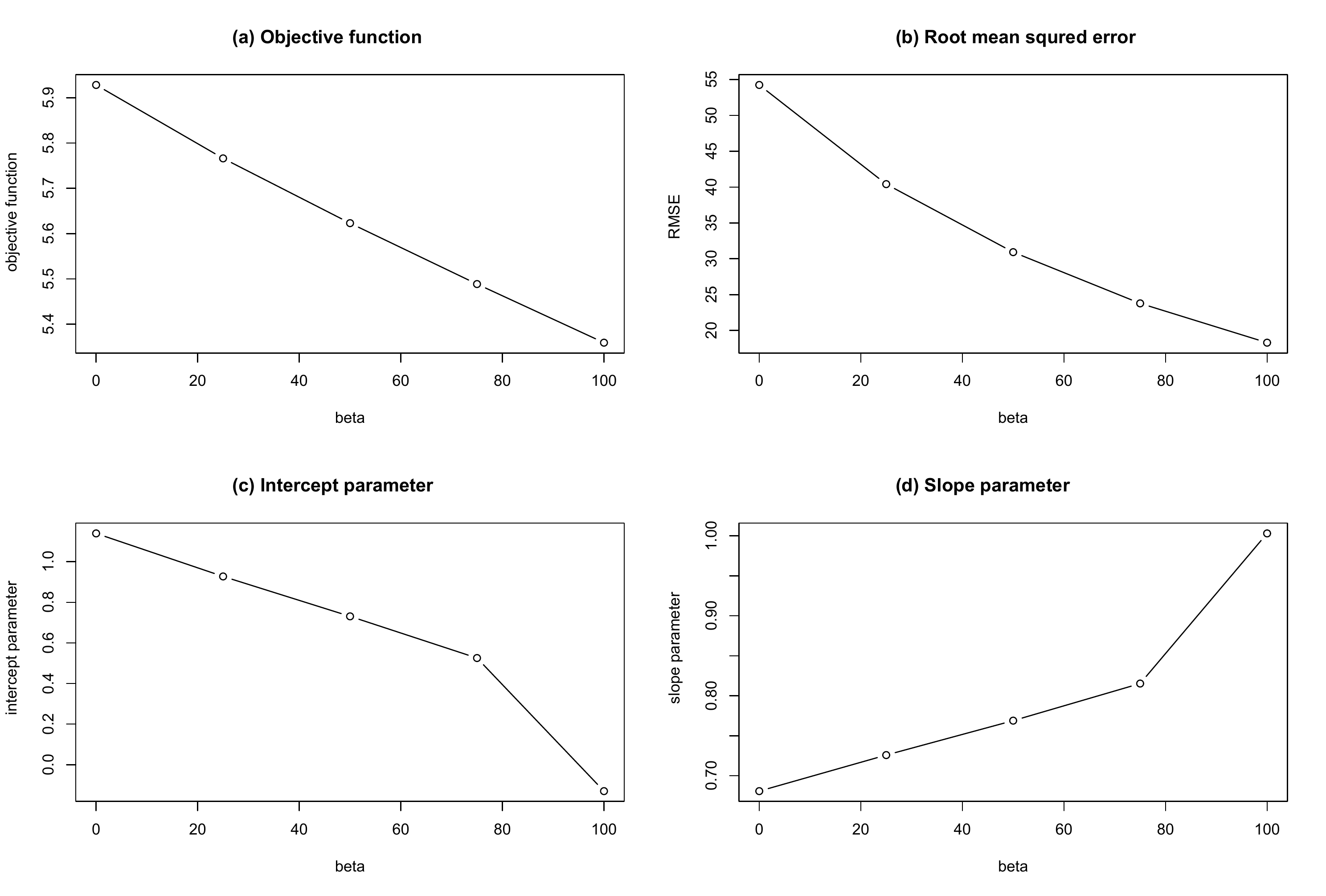}
\caption{The parameter $\beta$ dependences of the reconstruction for No-cut data with link constraints}
\label{fig:nocut_with_lc}       
\end{figure*}

\begin{figure*}
  \includegraphics[width=0.5\textwidth]{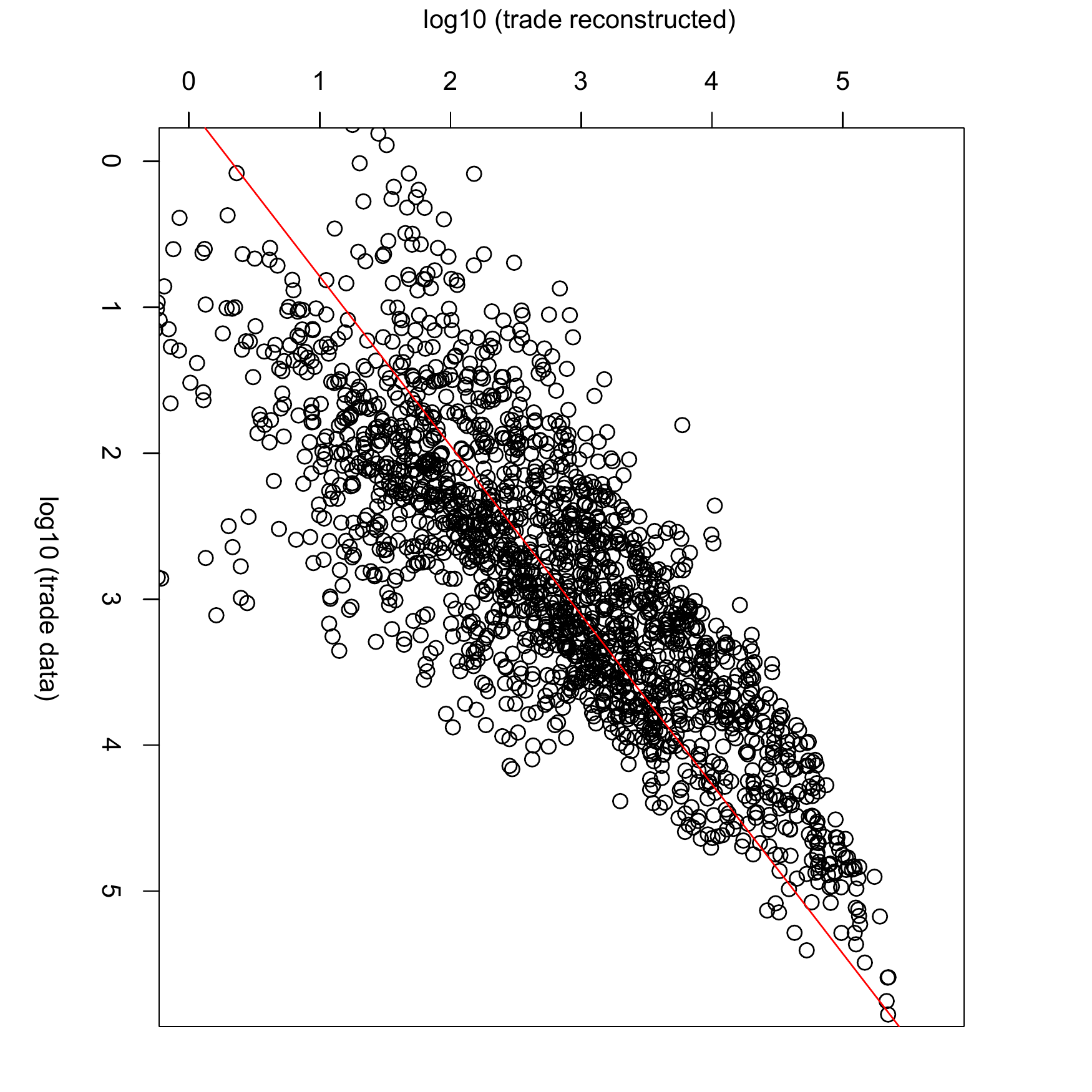}
\caption{Relationship between the trade data and the reconstructed trade for No-cut data with link constraints with $\beta=100$}
\label{fig:Ridge5_nocut}       
\end{figure*}

Second, we reconstruct the trade network $t_{ij}^{(reconst)}$ with the link constraints.
The parameter $\beta$ dependences of the reconstruction are shown in Fig. \ref{fig:nocut_with_lc}.
The observed parameter $\beta$ dependences are similar to the reconstruction without the link constraints.
Note that the intercept $b$ and the slope $a$ are approximately equal to 0 and 1, respectively, when $\beta$ is equal to $100$.
Figure \ref{fig:Ridge5_nocut} shows the relationship between the trade data and the reconstructed trade for No-cut data with link constraints with $\beta=100$.

\begin{figure*}
  \includegraphics[width=1.0\textwidth]{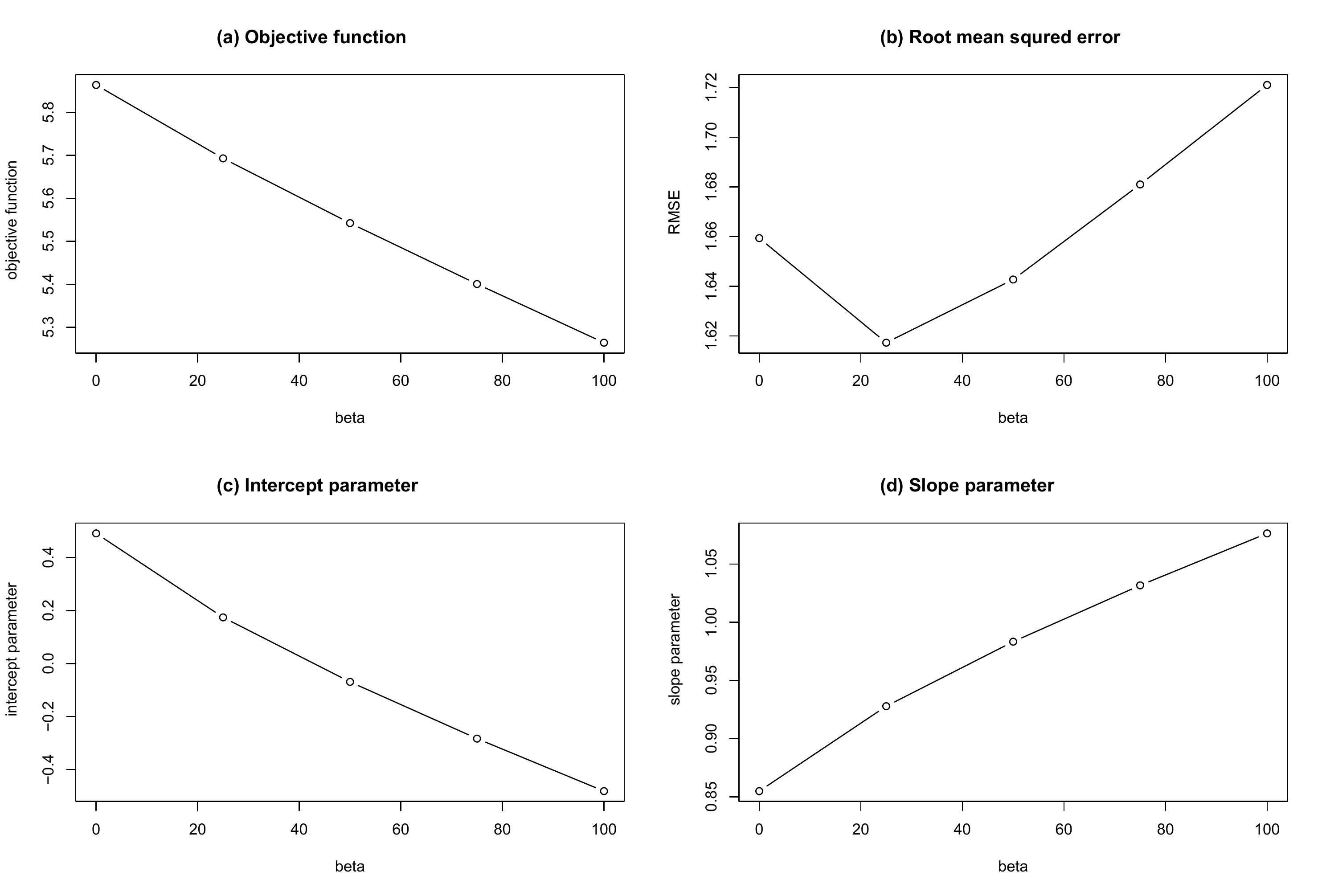}
\caption{The parameter $\beta$ dependences of the reconstruction for 2Q-cut data without link constraints}
\label{fig:2Qcut_wo_lc}       
\end{figure*}

Third, we reconstruct the trade network $t_{ij}^{(reconst)}$ without the link constraints for 2Q-cut data.
The parameter $\beta$ dependences of the reconstruction are shown in Fig. \ref{fig:2Qcut_wo_lc}.
The observed parameter $\beta$ dependences are similar to the reconstruction of No-cut data.
We notice that the intercept $b$ and the slope $a$ are approximately equal to 0 and 1, respectively when $\beta$ is equal to $50$.

\begin{figure*}
  \includegraphics[width=1.0\textwidth]{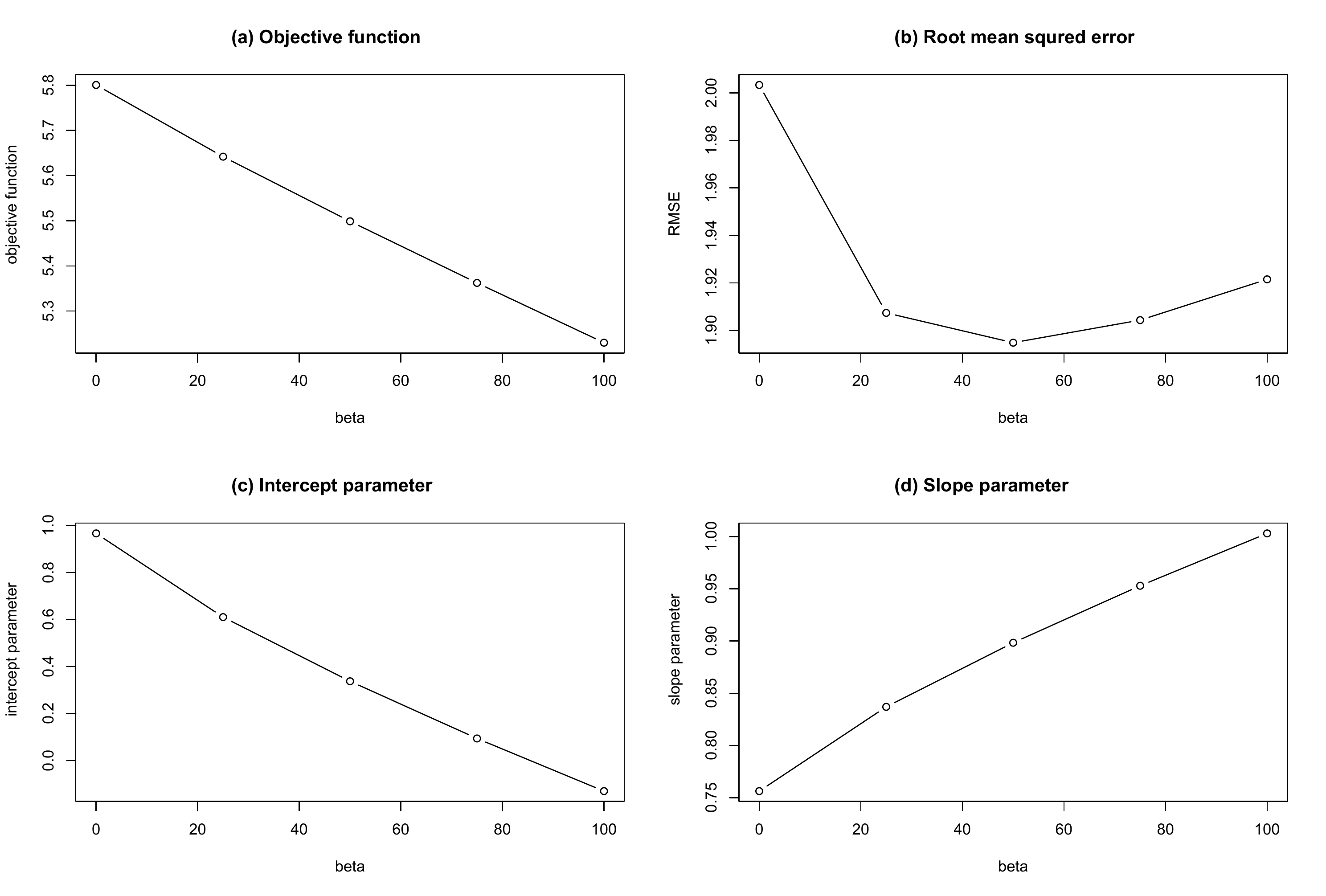}
\caption{The parameter $\beta$ dependences of the reconstruction for 2Q-cut data with link constraints}
\label{fig:2Qcut_with_lc}       
\end{figure*}

\begin{figure*}
  \includegraphics[width=0.5\textwidth]{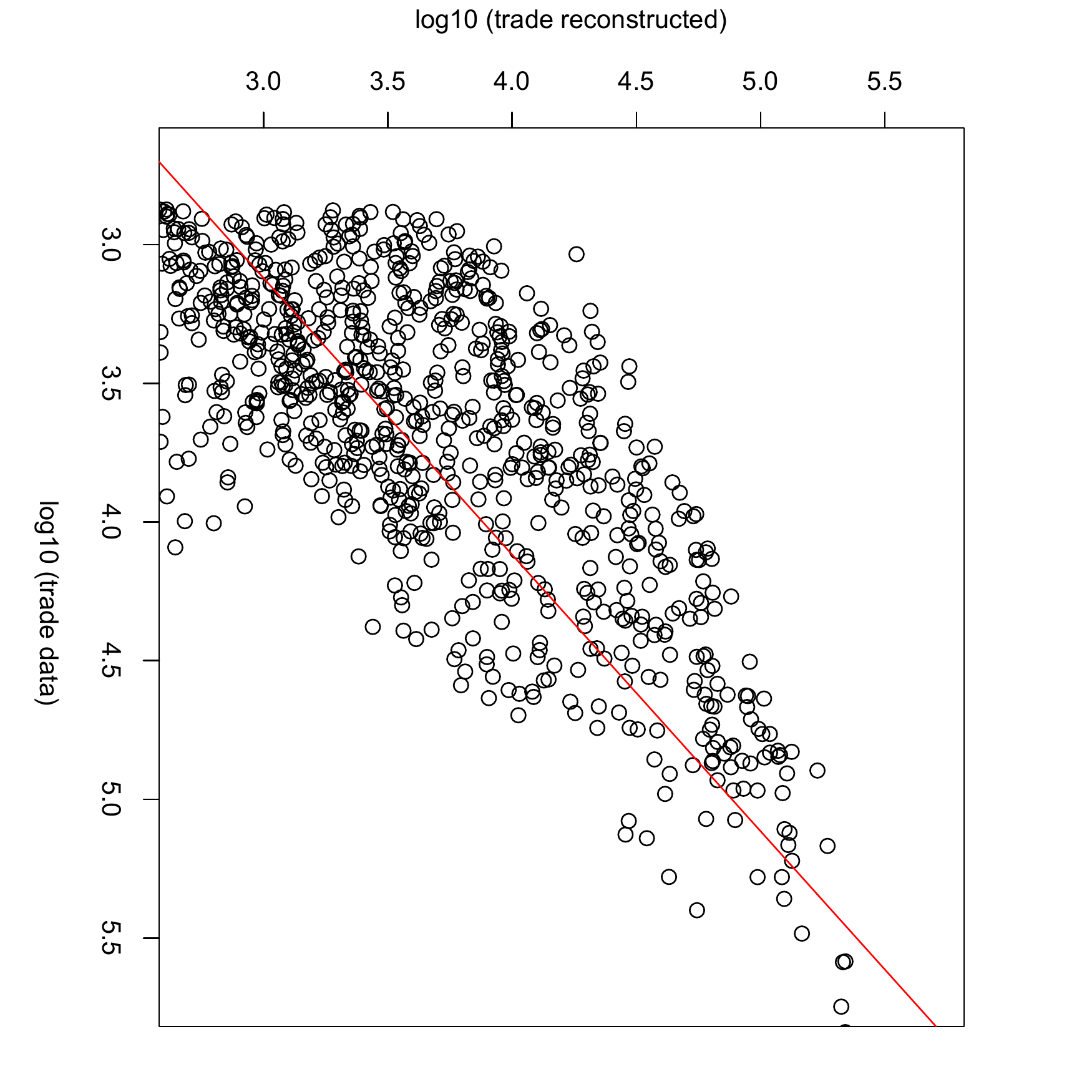}
\caption{Relationship between the trade data and the reconstructed trade for 2Q-cut data with link constraints with $\beta=100$}
\label{fig:Ridge5_2Qcut}       
\end{figure*}

\begin{figure*}
  \includegraphics[width=0.5\textwidth]{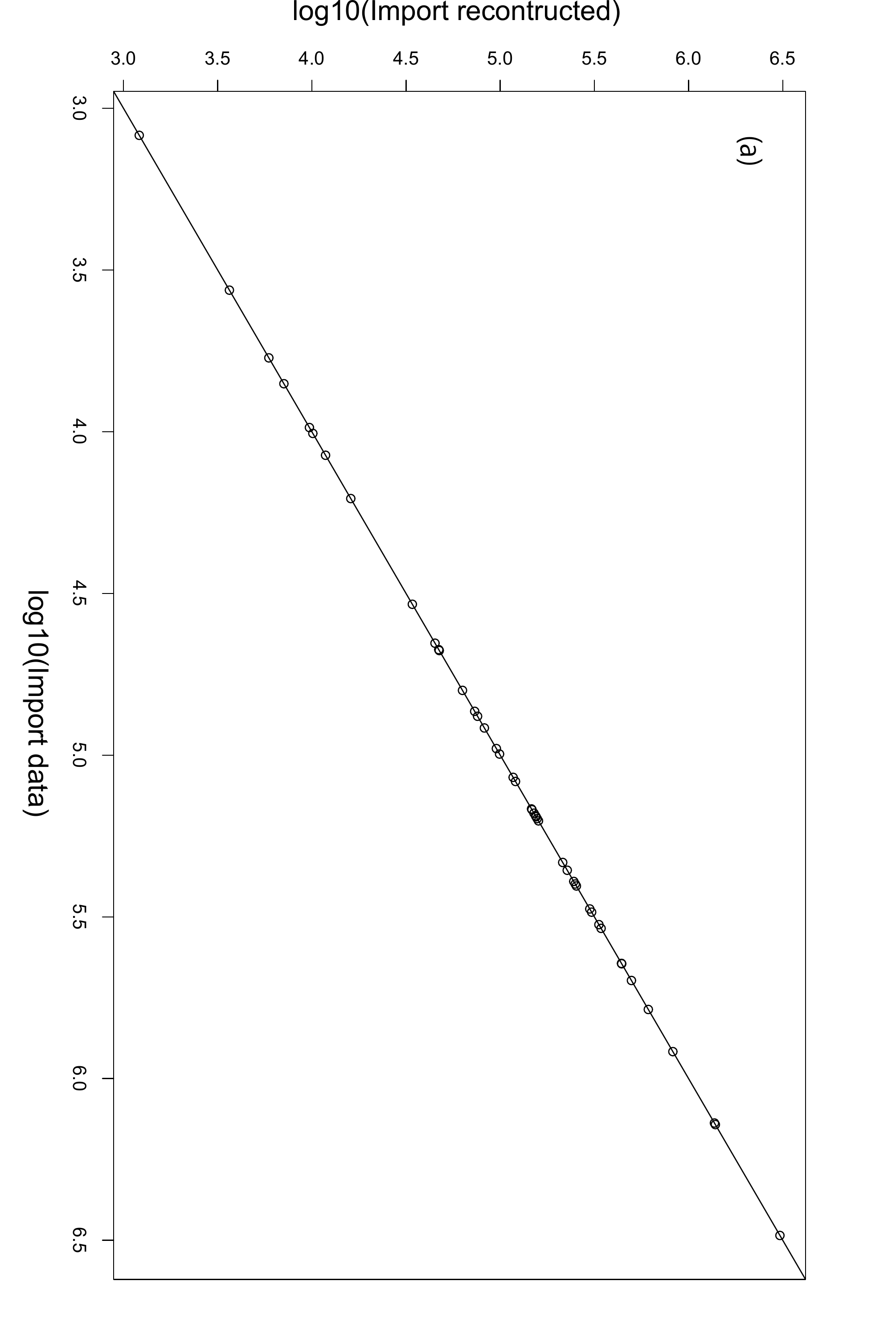}
  \includegraphics[width=0.5\textwidth]{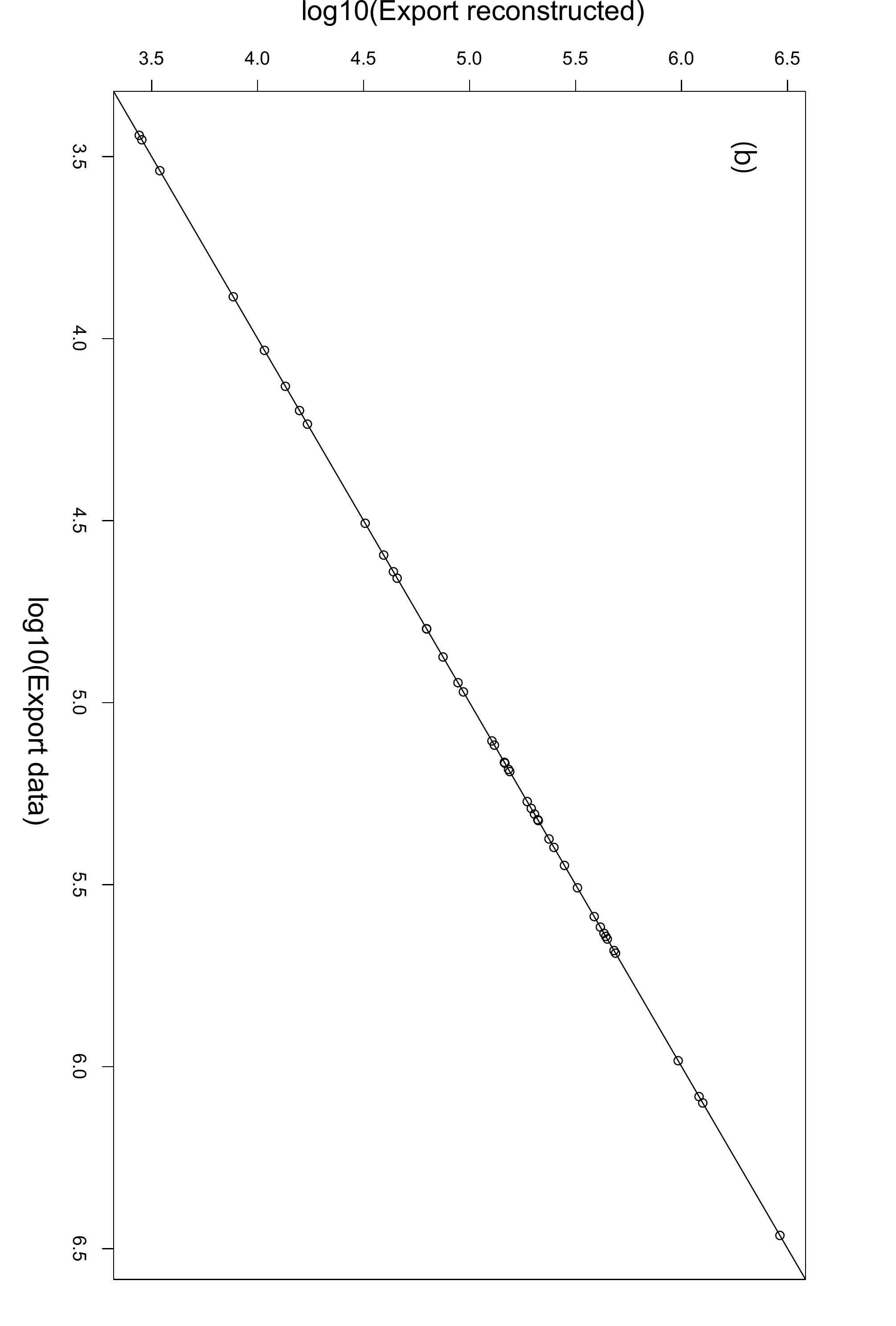}
\caption{Merginal distribution of the reconstraction for 2Q-cut data with link constraints}
\label{fig:2Qcut_merginal_with_lc}       
\end{figure*}

Fourth, we reconstruct the trade network $t_{ij}^{(reconst)}$ with the link constraints from 2Q-cut data.
The parameter $\beta$ dependences of the reconstruction are shown in Fig. \ref{fig:2Qcut_with_lc}.
The observed parameter $\beta$ dependences are similar to the reconstruction of 2Q-cut data without the link constraints.
Note that the intercept $b$ and the slope $a$ are approximately equal to 0 and 1, respectively, when $\beta$ is equal to $100$.
Figure \ref{fig:Ridge5_2Qcut} shows the relationship between the trade data and the reconstructed trade for 2Q-cut data with link constraints with $\beta=100$.
Reproduction of the marginal distribution of the reconstruction for 2Q-cut data with link constraints is confirmed in Fig. \ref{fig:2Qcut_merginal_with_lc}. In panel (a), the horizontal and vertical axes indicate the aggregated import data of each country and the aggregated export calculated from the reconstructed network, respectively.
Panel (b) is the same, but it is about aggregated exports.

The Network densities calculated for the four reconstructed networks are shown in Table \ref{table:density}.
The Network densities of the trade data are $1.0$ and $0.488$ for No-cut data and 2Q-cut data, respectively.
The reconstruction without the link constraints gave the network density equal to $1.0$ for both No-cut data and 2Q-cut data.
On the other hand, when the link constraints are used, the reconstruction gave the accurate value $0.488$ for 2Q-cut data.
This means that without using the link constraints, the Ridge Entropy Maximization does not reproduce the observed sparsity of the trade network.
\begin{table}[hbtp]
  \caption{Network density}
  \label{table:density}
  \centering
  \begin{tabular}{lcc}
    \hline
    item  & No-cut   & 2Q-cut  \\
    \hline \hline
    trade data                     & 1.0 & 0.488 \\
    reconstraction w/o link const.  & 1.0 & 1.0     \\
    reconstraction with link const.& 1.0 & 0.488  \\
    \hline
  \end{tabular}
\end{table}

It is, however, possible to obtain the desired network density by truncating small weighted links of the network.
For 2Q-cut data, if we truncate links, those weights are smaller than $0.24 \%$ of the maximum weighted link of the network,
the network density is reproduced so that the density is equal to $0.488$.

\begin{figure*}
  \includegraphics[width=0.5\textwidth]{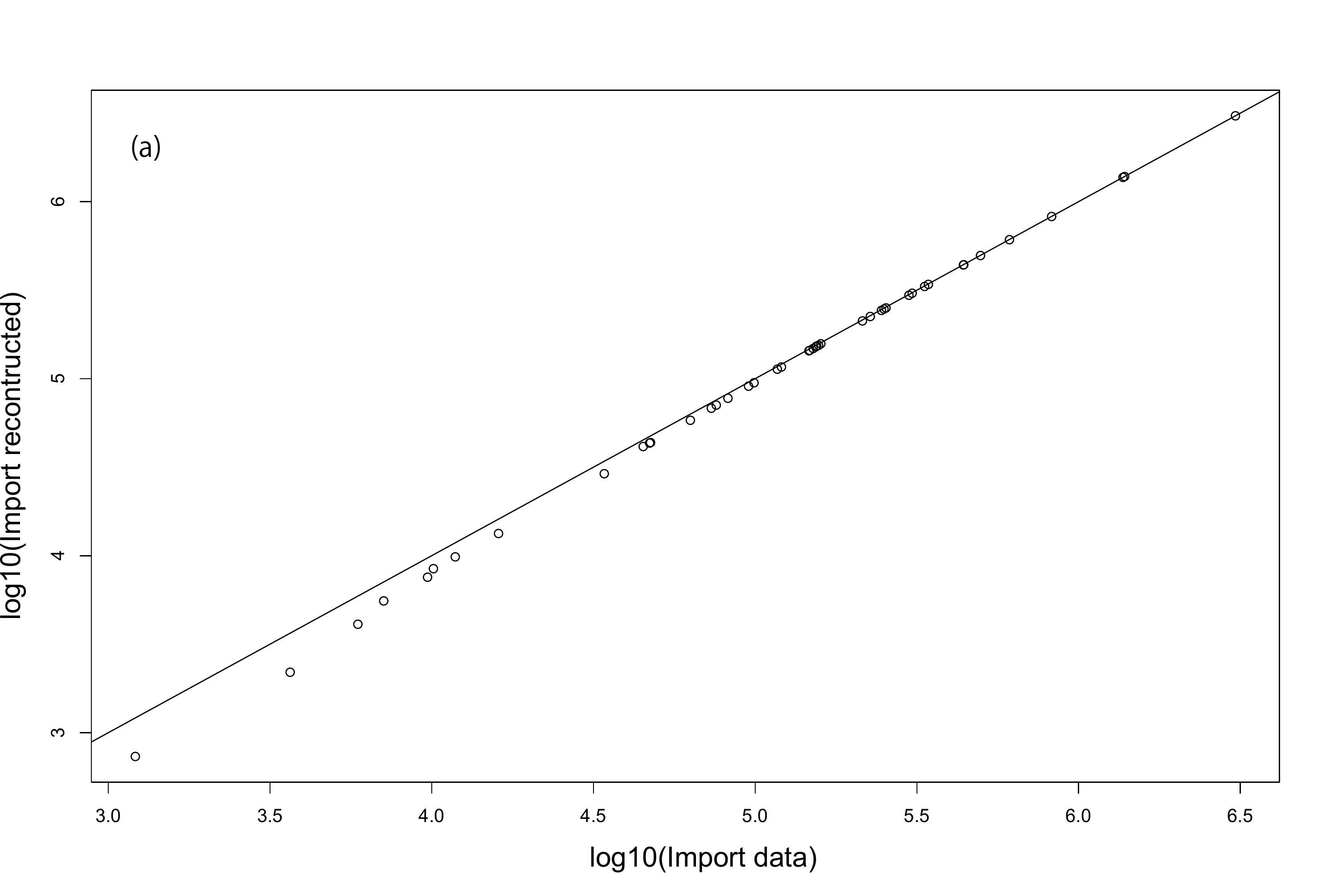}
  \includegraphics[width=0.5\textwidth]{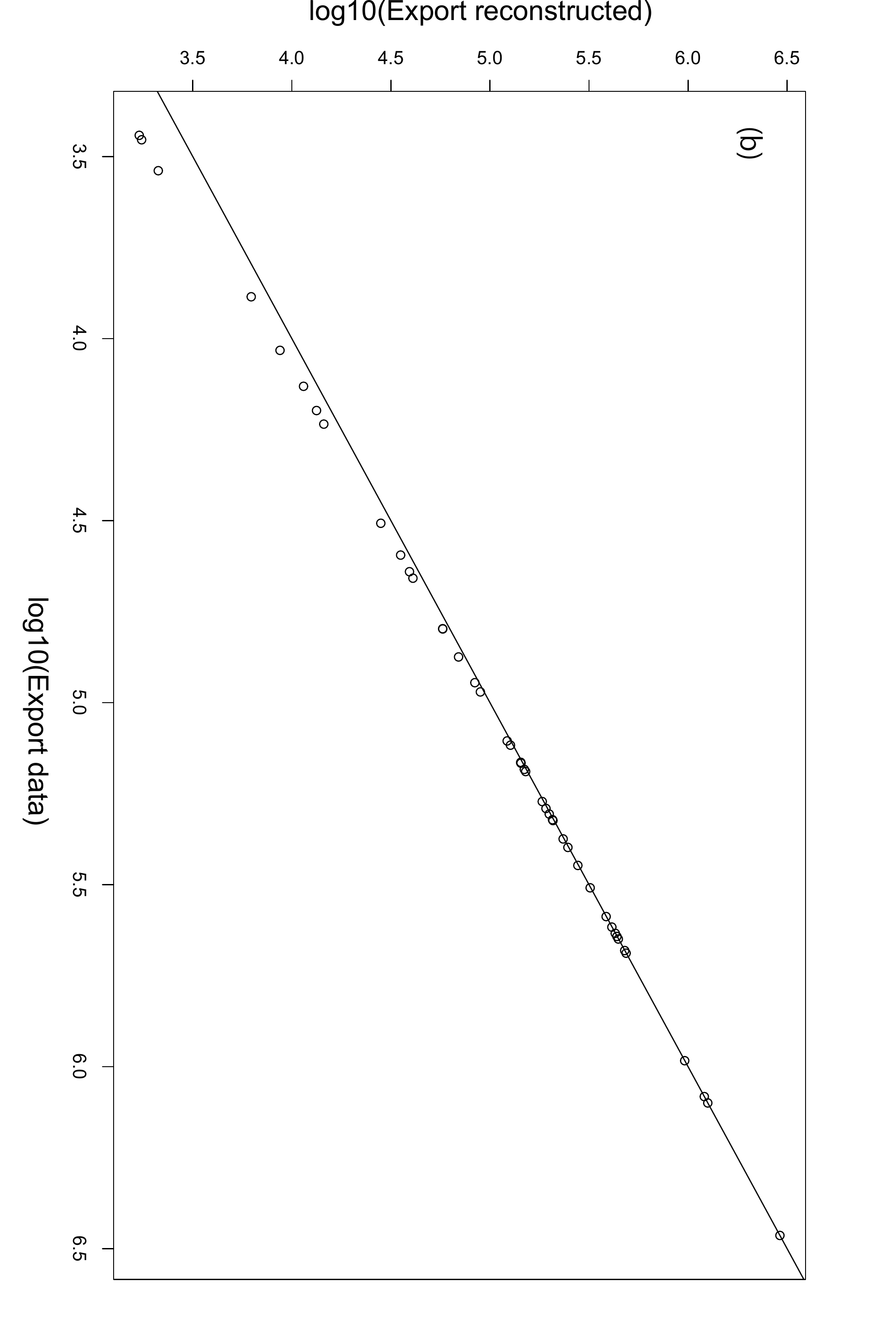}
\caption{Merginal distribution 2Q-cut without link constraints by trancating small weighted links. We notice that the aggregated import and export calculated from the reconstructed network are slightly lower than the actual data.}
\label{fig:2Qcut_merginal_wo_lc_with_trancation}       
\end{figure*}

The marginal distribution of the reconstruction for 2Q-cut data with link constraints by truncating small weighted links of the network is shown in Fig. \ref{fig:2Qcut_merginal_wo_lc_with_trancation}.
In panel (a), the horizontal and vertical axes indicate the aggregated import data of each country and the aggregated import calculated from the reconstructed network, respectively.
Panel (b) is the same, but it is about aggregated exports.
Note that the aggregated import and export calculated from the reconstructed network are slightly lower than the actual data in Fig. \ref{fig:2Qcut_merginal_wo_lc_with_trancation}.
We can say that the deviation of the marginal distribution is small when the truncation of small weighted links is applied to obtain the observed sparsity of the network.

\section*{Appendix B}
\label{sec:8}

Two large communities were detected in 2000, 2007, and 2015, which showed relatively large values of modularity. In these communities, we listed the 10 financial institutions with the highest PageRank as major nodes. The major nodes of the first and second communities in 2000 are shown in Table \ref{table:MN2000Comm1} and Table \ref{table:MN2000Comm2}; the major nodes of the first and second communities in 2007 are shown in Table \ref{table:MN2007Comm1} and Table \ref{table:MN2007Comm2}; the major nodes of the first and second communities in 2015 are shown in Table \ref{table:MN2015Comm1} and Table \ref{table:MN2015Comm2}.

\begin{table}[hbtp]
  \caption{Major Nodes of Community 1 in 2000}
  \label{table:MN2000Comm1}
  \centering
  \begin{tabular}{lcccc}
    \hline
    bank name &  bank category & in-degree & out-degree & PageRank \\
    \hline \hline
Mizuho Bank & major commercial  & 94 & 23 & 0.100005 \\
MUFG Bank & major commercial  & 93 & 56 & 0.065976 \\
Aozora Bank & major commercial  & 93 & 18 & 0.063958 \\
Snisei Bank & major commercial  & 84 & 20 & 0.059184 \\
Kirayaka Holdings & second-tier regional  & 63 & 21 & 0.055473 \\
Mizuho Trust and Banking & trust  & 65 & 0 & 0.046078 \\
Sumitomo Mitsui Trust Bank & trust  & 65 & 25 & 0.046078 \\
Nishi Nippon City Bank & second-tier regional  & 52 & 1 & 0.043203 \\
Resona Bank & major commercial  & 56 & 37 & 0.042650 \\
Bank of Fukuoka & leading regional  & 7 & 9 & 0.014147 \\
    \hline
  \end{tabular}
\end{table}
\begin{table}[hbtp]
  \caption{Major Nodes of Community 2 in 2000}
  \label{table:MN2000Comm2}
  \centering
  \begin{tabular}{lcccc}
    \hline
    bank name &  bank category & in-degree & out-degree & PageRank \\
    \hline \hline
Shizuoka Bank & leading regional  & 34 & 9 & 0.024182 \\
Bank of Yokohama & leading regional  & 33 & 9 & 0.023255 \\
Joyo Bank & leading regional  & 30 & 10 & 0.021767 \\
Kagoshima Bank & leading regional  & 22 & 9 & 0.019465 \\
Gunma Bank & leading regional  & 20 & 5 & 0.019044 \\
San In Godo BANK & leading regional  & 20 & 8 & 0.019044 \\
Chugoku Bank & leading regional  & 20 & 9 & 0.019044 \\
Hachijuni Bank & leading regional  & 14 & 9 & 0.017887 \\
Ogaki Kyoritsu Bank & leading regional  & 14 & 9 & 0.017887 \\
Hyakujushi Bank & leading regional  & 13 & 5 & 0.017782 \\
    \hline
  \end{tabular}
\end{table}

\begin{table}[hbtp]
  \caption{Major Nodes of Community 1 in 2007}
  \label{table:MN2007Comm1}
  \centering
  \begin{tabular}{lcccc}
    \hline
    bank name &  bank category & in-degree & out-degree & PageRank \\
    \hline \hline
Shimizu Bank & leading regional  & 39 & 8 & 0.052739 \\
Nanto Bank & leading regional  & 39 & 10 & 0.052739 \\
Daishi Bank & leading regional  & 32 & 11 & 0.048186 \\
Shonai Bank & leading regional  & 32 & 10 & 0.048186 \\
MU Trust and Banking & trust  & 6 & 14 & 0.008441 \\
Aozora Bank & major commercial  & 6 & 15 & 0.008441 \\
Toho Bank & leading regional  & 6 & 10 & 0.004501 \\
Sumitomo Mitsui Trust Bank & trust  & 4 & 14 & 0.004176 \\
Bank of Kyoto & leading regional  & 4 & 13 & 0.004176 \\
Tomato Bank & second-tier regional  & 2 & 7 & 0.004118 \\

    \hline
  \end{tabular}
\end{table}
\begin{table}[hbtp]
  \caption{Major Nodes of Community 2 in 2007}
  \label{table:MN2007Comm2}
  \centering
  \begin{tabular}{lcccc}
    \hline
    bank name &  bank category & in-degree & out-degree & PageRank \\
    \hline \hline
MUFG Bank & major commercial  & 85 & 63 & 0.078086 \\
Resona Bank & major commercial  & 85 & 43 & 0.077606 \\
Mizuho Bank & major commercial  & 85 & 16 & 0.075147 \\
Kagawa Bank & second-tier regional  & 67 & 7 & 0.054281 \\
Snisei Bank & major commercial  & 57 & 13 & 0.046089 \\
Bank of Minami Nippon & second-tier regional  & 62 & 8 & 0.044499 \\
Towa Bank & second-tier regional  & 62 & 7 & 0.044499 \\
Momiji Bank & second-tier regional  & 62 & 8 & 0.044499 \\
Taiko Bank & second-tier regional  & 62 & 7 & 0.044499 \\
Kirayaka Holdings & second-tier regional  & 46 & 8 & 0.034558 \\
    \hline
  \end{tabular}
\end{table}

\begin{table}[hbtp]
  \caption{Major Nodes of Community 1 in 2015}
  \label{table:MN2015Comm1}
  \centering
  \begin{tabular}{lcccc}
    \hline
    bank name &  bank category & in-degree & out-degree & PageRank \\
    \hline \hline
Bank of Minami Nippon & second-tier regional  & 67 & 1 & 0.071720 \\
Taiko Bank & second-tier regional  & 67 & 1 & 0.071720 \\
Mizuho Bank & major commercial  & 69 & 18 & 0.062541 \\
MUFG Bank & major commercial  & 69 & 17 & 0.062376 \\
Resona Bank & major commercial  & 69 & 10 & 0.060364 \\
Momiji Bank & second-tier regional  & 51 & 8 & 0.054483 \\
Sumitomo Mitsui Trust Bank & trust  & 67 & 15 & 0.053086 \\
MU Trust and Banking & trust  & 67 & 12 & 0.053086 \\
Mizuho Trust and Banking & trust  & 67 & 8 & 0.053086 \\
Kirayaka Holdings & second-tier regional  & 16 & 9 & 0.018744 \\
    \hline
  \end{tabular}
\end{table}
\begin{table}[hbtp]
  \caption{Major Nodes of Community 2 in 2015}
  \label{table:MN2015Comm2}
  \centering
  \begin{tabular}{lcccc}
    \hline
    bank name &  bank category & in-degree & out-degree & PageRank \\
    \hline \hline
Fukui Bank & leading regional  & 39 & 9 & 0.174540 \\
Bank of Yokohama & leading regional  & 27 & 13 & 0.038477 \\
Snisei Bank & major commercial  & 21 & 13 & 0.018963 \\
Chugoku Bank & leading regional  & 17 & 9 & 0.018485 \\
Hokkoku Bank & leading regional  & 14 & 2 & 0.016412 \\
Aozora Bank & major commercial  & 16 & 11 & 0.014013 \\
Gunma Bank & leading regional  & 7 & 11 & 0.010706 \\
Shizuoka Bank & leading regional  & 5 & 13 & 0.004579 \\
Chukyo Bank & second-tier regional  & 6 & 1 & 0.001916 \\
Tomato Bank & second-tier regional  & 6 & 1 & 0.001916 \\
    \hline
  \end{tabular}
\end{table}

\begin{acknowledgements}
This study was funded by the Ministry of Education, Science, Sports, and Culture (Grants-in-Aid for Scientific Research (B), Grant No. 17KT0034.
\end{acknowledgements}

%
 \section*{Conflict of interest}
On behalf of all authors, the corresponding author states that there is no conflict of interest.



\end{document}